\documentclass[preprint]{aastex}

\begin{document}

\title{Exploring Halo Substructure with Giant Stars: II. Mapping the
  Extended Structure of the Carina Dwarf Spheroidal Galaxy }

\author{Steven R. Majewski\altaffilmark{1,2,3,4}, 
James C. Ostheimer\altaffilmark{1}, 
Richard J. Patterson\altaffilmark{1},
William E. Kunkel\altaffilmark{5},
Kathryn V. Johnston\altaffilmark{6}, and
Doug Geisler\altaffilmark{7}}

\altaffiltext{1}{Dept. of Astronomy, University of Virginia,
Charlottesville, VA, 22903-0818 (srm4n@didjeridu.astro.virginia.edu, 
jco9w@virginia.edu,
ricky@virginia.edu)}

\altaffiltext{2}{Visiting Associate, The Observatories of the Carnegie
Institution of Washington, 813 Santa Barbara Street, Pasadena, CA 91101}

\altaffiltext{3}{David and Lucile Packard Foundation Fellow}

\altaffiltext{4}{Cottrell Scholar of The Research Corporation}

\altaffiltext{5}{Las Campanas Observatory, Carnegie Institution of
Washington, Casilla 601, La Serena, Chile
(skunk@roses.ctio.noao.edu)}

\altaffiltext{6}{Dept. of Astronomy, Wesleyan University, Middletown, CT 06459-0123 (kvj@astro.wesleyan.edu)}

\altaffiltext{7}{Department of Physics, Universidad de Concepci\'on, Concepci\'on, Chile
(doug@stars.cfm.udec.cl)}

\begin{abstract}
  As part of our survey for substructure in the Milky Way halo as traced
  by giant stars, and to look for tidal stellar debris in the halo
  viable for measurement of the Galactic mass potential with the Space
  Interferometry Mission (SIM), we explore the distribution of stars
  beyond the nominal tidal radius of, but still associated with, the
  Carina dwarf spheroidal galaxy. We make use of the photometric
  technique described in Majewski et al.\ (1999b, AJ, submitted) to
  identify giant star candidates at the distance and metallicity of the
  Carina dwarf spheroidal across the entire extent of a photometric
  survey covering some 2.2 deg$^2$ on and around Carina.  These
  Carina-associated giant candidates are identified by a combination of
  (1) their $M-DDO51$ colors, which are a measure of both surface
  gravity and metallicity at given $M-T_2$ colors, and (2) by locations
  in the color-magnitude diagram commensurate with the Carina red giant
  branch in the core of the galaxy.  The density distribution of the
  extratidal giant candidates bears resemblance to the outer isopleths
  of Carina presented by Irwin \& Hatzidimitriou (1995, \mnras, 277,
  1354).  However, in contrast to previous, {\it statistical}
  star-counting approaches, we can pinpoint {\it actual},
  remotely-situated Carina stars individually.  Because we can exclude
  the foreground veil of dwarf stars, our approach allows greater
  sensitivity and the ability to map the detailed two-dimensional
  distribution of extended Carina populations to much larger radii,
  while utilizing smaller aperture telescopes, than other techniques.
  Moreover, we identify candidate lists of widely displaced
  Carina-associated stars bright enough for spectroscopic studies of
  large-scale dynamical and metallicity properties of the system, and
  for astrometric study by SIM. We obtained spectroscopy for three such
  ``extratidal'' stars and from their radial velocities conclude that
  all three are associated with Carina.

  While a single King profile matches our derived Carina core density
  profile, we confirm previous claims for a break in the density
  fall-off at about 20 arcmin.  Beyond this radius, a more gradual
  fall-off as $r^{-\gamma}$, with $1 < \gamma < 2$, to $r \ge 80$
  arcmin, is found. If the existence of density profile breaks is a
  signature of the predominance of unbound stars, and if we adopt the
  nominal tidal radius of 28 arcmin previously found for Carina, then it
  would appear that we have identified a substantial extratidal
  population from Carina.  If these $r \gtrsim 20$ arcmin stars are
  truly now unbound from the galaxy, we estimate from the relative
  stellar density distribution a fractional destruction rate for Carina
  from tidal stripping of order $\begin{pmatrix} {df \over dt}\end{pmatrix} =
  0.27$ Gyr$^{-1}$.  This is among the highest rates expected for the
  Milky Way dwarf spheroidals apart from Sagittarius.  The existence of
  such extended populations of Carina-associated stars may have
  important implications for the existence of large dark matter contents
  in dwarf spheroidals, as well as for the evolution of the Milky Way
  halo.
  
  Finally, we find that the ``background density'' of what are likely to
  be predominantly random, metal-poor halo field giants maintains a
  rather flat count-magnitude relation out to the distance of Carina, in
  keeping with $R^{-3}$ density laws for the Galactic halo.
\end{abstract}


\section{Introduction}
It is becoming increasingly clear that dwarf spheroidal galaxies are not
the simple systems they were once thought to be.  A great deal of recent
work has exposed the complicated star formation histories in many of
these small systems in the Local Group (see summaries by Grebel 1997,
1998, Mateo 1998).  That a system like Carina, which has experienced at
least three major episodes of star formation in its lifetime
(Smecker-Hane et al.\ 1994, 1996, Mighell 1997, Hurley-Keller et al.\ 
1998), can retain gas to fuel repeated bursts appears to contradict old notions of
these low luminosity systems having fragile, fluffy mass potentials
(see, e.g., Dekel \& Silk 1986, Burkert \& Ruiz-Lapuente 1997, MacLow \&
Ferrara 1999).

Indeed, dynamical studies of the mass-to-light, $M/L$, ratios of most of
the Galactic dwarf spheroidals (e.g., Aaronson 1983, Seitzer \& Frogel
1985, Aaronson \& Olszewski 1987, Pryor \& Kormendy 1990, Mateo et al.\
1993, Suntzeff et al.\ 1993, Hargreaves et al.\ 1994, Vogt et al.\ 1995,
Olszewski et al.\ 1996, Mateo et al.\ 1998b) imply large values,
approaching $M/L \sim 100$ or more in the systems with the least total
luminosity: Draco, Sextans and Ursa Minor (Olszewski, Aaronson \& Hill
1995 and references therein, Irwin \& Hatzidimitriou 1995, IH95
hereafter; Mateo 1998).  Clearly dwarf spheroidal galaxies are not just
larger versions of globular clusters, which have typical $M/L \sim 1-2$,
but are of a very different structural character. The structural
difference between globular clusters and dwarf spheroidals has often
been attributed to a large dark matter content in the latter.

However, the large dark matter interpretation of the large velocity
dispersions observed in dwarf spheroidals is subject to various
uncertainties and still debated.  There are assumptions (see Vogt et
al.\ 1995, Piatek \& Pryor 1995, IH95, Kleyna et al.\ 1999) incorporated
into typical $M/L$ determinations that remain to be verified, including
the assumption of isotropically distributed stellar orbits, that mass
follows the distribution of light in these systems, and even that normal
Newtonian gravity applies.\footnote{Milgrom (1983a,b) has proposed a
  form of Modified Newtonian Dynamics (MOND) that serves to reduce the
  implied $M/L$'s of bound stellar systems.}  Moreover, unresolved
binaries inflate derived velocity dispersions to some degree that may or
may not affect $M/L$ ratios significantly (Suntzeff et al.\ 1993,
Olszewski, Pryor \& Armandroff 1996, Hargreaves, Gilmore \& Annan 1996).

Perhaps the greatest uncertainty in the dark matter interpretation of
the velocity dispersion data is that of a virial equilibrium state for
the dwarf spheroidals.  The notion of dwarf galaxies in virial
equilibrium has been questioned by, e.g., Kuhn \& Miller (1989), who
attributed the high velocity dispersions to orbital resonance
``heating'' of the stars in satellites, with a resulting inflation of
the internal velocities.  While this particular model has been
controversial (Pryor 1999), the idea that passage of the satellites near
massive objects like the Galactic center or dark matter clumps in the
halo (Kroupa 1999) can affect the internal dynamics and outer structure
of these dwarf galaxies has a long history.

For example, Hodge \& Michie (1969) concluded that Galactic tides could
have an important effect on the outer structure of dwarf spheroidals and
suggested that Ursa Minor is presently in the throes of total
disruption.  Still, it is not clear to what extent tidal disruption can
perturb the inferred dark matter content.  For example, it is hard to
understand how tides could be affecting such distant dwarfs as Leo I and
Leo II, for which relatively large ($\sim 10$), $M/L$ are found (Vogt et
al.\ 1995).  In a numerical modeling study of the phenomenon of tidally
induced, large velocity dispersions in dwarf spheroidals, Piatek \&
Pryor (1995) concluded that Galactic tides cannot account for 
extraordinarily large $M/L$, though an inflation of $M/L$ to about 40
was possible.  Johnston, Sigurdsson \& Hernquist (1999b) similarly find
little influence on the core velocity dispersions for tidally disrupting
systems.  The main influence of tides on the dynamics of the dwarfs is
not to inflate central velocity dispersions, but rather to produce large
ordered motions that would resemble apparent systemic rotations.  Indeed, such
shearing motions have been observed in the Ursa Minor and Draco systems
by Hargreaves et al.\ (1994).

However, a contrasting point of view comes out of high resolution,
N-body studies by Klessen \& Kroupa (1998) of the dynamical evolution of
satellite galaxies in Milky Way-like gravitational potentials {\it
  followed until well after the disruption}. Klessen \& Kroupa find that
the debris of large satellite galaxies on orbits of eccentricities
greater than 0.41 and undergoing severe tidal disruption eventually
converge into stable remnants of about 1\% the original satellite mass.
When viewed along certain lines of sight, these remnants have properties
very similar to dwarf spheroidal galaxies, including, most
interestingly, velocity dispersions leading to {\it inferred} high
$M/L$'s {\it in spite of the fact that the remnants do not contain dark
  matter}.  The notion that some of the Milky Way dwarf spheroidals may
be tidal remnants has been discussed for several decades because of the
seemingly non-random alignments of the satellites around the Milky Way
(Kunkel 1979, Lynden-Bell 1982, Majewski 1994, Lynden-Bell \&
Lynden-Bell 1995, Palma, Majewski \& Johnston 1999).  While Klessen \&
Kroupa do not predict what the density profiles of their tidal remnants
would look like, their scenario predicts that their dSphs should exhibit
ordered radial velocity gradients across the galaxy, similar to those
discussed above.

These issues have been confused by the discovery around the majority of
the Milky Way dwarf spheroidals of ``breaks'' in the starcount profiles
where the character of the counts changes from a steeply falling King
profile to a much more gradual decline with radius (Eskridge 1988a,b;
IH95).  Such features are seen in simulations in which a satellite (with
mass tracing the light) is being stripped by the Milky Way's tidal field
(Oh, Lin \& Aarseth 1995, Piatek \& Pryor 1995, Johnston et al.\ 1999b),
and it is tempting to attribute such breaks in the radial profile to the
onset of an outer population of escaping stars.\footnote{ It is worth
  noting one interesting exception to this attribution.  In an earlier
  study of the extratidal phenomenon with van Agt's (1978) sample of
  extratidal stars around Sculptor, Innanen \& Papp (1979) concluded
  that stars outside the tidal radius could still be bound if on
  retrograde orbits about the satellite.  Stars on such orbits can
  resist tidal stripping by the Milky Way.}  On the other hand,
arguments have been made that the existence of tidal tails is evidence
against the presence of significant dark matter halos in dwarf
spheroidals (Moore 1996, Burkert 1997).  If the satellite contained
sufficient dark matter, then the point at which tidal effects became
important would lie well beyond the ``tidal radius'' found from King
model fits to the luminous matter.  As Moore (1996) concludes: ``...even
modest amounts of dark matter will be very effective at containing the
visible stars and halting the production of tidal tails.''

Whatever the solution to these contradictory and confusing issues, it is
certain that more observational handles on the problem would help with
clearing the theoretical hurdles.  For example, if the outer populations
of the dwarf spheroidals could be mapped well past the break radius, it
may become evident whether they evolve into obvious tidal tails.
Measuring the velocities of stars beyond the break would also provide a
great advantage to discriminating between models.  For example, evidence
for shearing motions, as described above, would support tidal tail
models, while isotropic velocity dispersions would support the notion
that the beyond-the-break populations are bound.  Both mapping dwarf
galaxies to large radii and obtaining velocities of well separated, but
associated stars are among the goals of the present program, which has
an overall aim to employ various strategies to uncover tidal debris from
disrupting satellite galaxies.

In this contribution we present the first results of a search for widely
extended, beyond-break-radius stars associated with nearby dwarf
galaxies, with a focus on a survey around the Carina dwarf spheroidal.
Carina was discovered from UK Schmidt plates by Cannon, Hawarden, \&
Tritton (1977). Demers, Beland, \& Kunkel (1983) determined the
following structural parameters (from star counts off of prints of a
CTIO 4-m plate) out to a radius of $\sim38$ arcmin along the semimajor
axis: an ellipticity of 0.4 at a position angle of 75$^{\circ}$, and a
tidal radius of 33 arcmin.  IH95 examined Carina's structure via star
counts out to a radius of $\sim40$ arcmin using APM scans of Schmidt
plates and found similar structural parameters: an ellipticity of 0.33
at a position angle of 65$^{\circ}$, and a tidal radius of 28 arcmin.
Most significantly, IH95 found a clear break in the radial starcount
profile and suggested the existence of an apparent extratidal
population.  Kuhn, Smith \& Hawley (1996) noted a spatially extended RR
Lyrae distribution for Carina (though all of their RR Lyrae were
interior to the IH95 tidal radius), and with a more sophisticated
starcount analysis gave even more compelling evidence for an extratidal
Carina population extending some four tidal radii along the major axis,
twice as far out as measured by IH95.  From theoretical considerations
of the observed structural parameters of Galactic dwarfs in IH95,
Johnston et al.\ (1999b) designated Carina as one of the likely Galactic
dwarf spheroidals with among the highest current fractional destruction
rates.  Given these various encouraging indicators, Carina is a
promising first candidate to test our search technique for true tidal
debris among the Galactic dwarf spheroidals.

We have reported elsewhere (Majewski 1999, Majewski et al.\ 1999c)
preliminary results from our search for tidal debris around the
Magellanic Clouds.  Our overall strategy for finding extratidal stars
differs from previous efforts in that we specifically target {\it giant
  stars} associated with the dwarf galaxies. This allows us to cover
large areas of the sky efficiently with small telescopes, as we do not
require deep imaging: Only the top several magnitudes of the red giant
branch (RGB) of each galaxy are sought.  The giant stars are identified
photometrically using the three filter, Washington $M, T_2 + DDO51$
system described in Majewski et al.\ (1999b; ``Paper I'' hereafter).
While perhaps more prone to small number statistics because of more restricted
sample sizes, our technique confers certain advantages over the deep
imaging, ``CMD-differencing'' strategies employed by, for example, IH95,
Kuhn et al.\ (1996), Grillmair et al.\ (1995; see also Grillmair 1998),
and others, that depend on uncovering {\it statistical} excesses of
starcounts (either total starcounts or counts in particular regions of
color-magnitude space).  The latter type studies are especially
sensitive to the zero-point level of, and variations in, the stellar
starcount background (see, e.g., the discussions on this point in IH95).
On the other hand, our goal is to pinpoint {\it actual} extratidal
candidates individually (not statistically) by their signal as giant
stars with properties expected for giants associated with the dwarf
galaxy.  By weeding out the overwhelming foreground curtain of dwarf
stars, our approach is not only less susceptible to background
subtraction problems and thereby capable of probing to larger radii more
easily, but we also generate candidate lists of {\it bona fide}
extratidal giants that are bright enough for spectroscopic verification
and study.  Thus, our approach fulfills the above stated strategies of
mapping to large radii as well as providing good candidates for
spectroscopy to do dynamical tests.

A campaign of targeted searches for extratidal stars and tidal tails
around Galactic dwarf galaxies is of interest for numerous reasons.
First, it is important to understand the extended structure of the dwarf
spheroidals as leverage on the dark matter issues outlined above.
Indeed, because we identify specific ``extratidal'' targets suitable for
radial velocity measurement, we hope to be able to test directly whether
they are bound to the dwarfs or not on the basis of the differences in
the expected dynamical signatures.  If unbound, we may then proceed to
test the various models (e.g., Oh et al.\ 1995, Piatek \& Pryor 1995,
Hamlin 1997, Johnston 1998, Johnston et al.\ 1999b) of dwarf spheroidal
disruption that make specific predictions of the velocity
characteristics of these stars.

Second, the discovery of substantial tidal tails associated with
Galactic satellites would provide the opportunity to measure the shape
and size of the Galactic mass potential with unprecedented accuracy
(Johnston et al.\ 1999c).  An advantage conferred by our survey approach
is that we can identify actual tidally-stripped stars that are viable
candidates (i.e., bright enough) for proper motion measurement via the
Space Interferometry Mission; a sample of some 100 such stars along a
tidal tail with fully measured space velocities (to 10 km s$^{-1}$
accuracy) can yield a measure of the mass of the Milky Way to a few
percent accuracy in the region that the satellite's orbit explores.
Hence, with several such tidal tails we could map the shape and size of
the Milky Way with unprecedented accuracy.

Finally, the possibility of ongoing destruction of globular clusters and
satellite galaxies has great bearing on the understanding of the
formation and evolution of our own Milky Way.  It is of interest to know
what contribution is made to the Galactic halo from the destruction of
satellites and accretion of their remains.  The substantial ongoing
contribution to the Milky Way of stellar and cluster debris from the
Sagittarius dwarf galaxy, with a tidal tail now mapped to some
40$^{\circ}$ from the galaxy core (Mateo, Olszewski \& Morrison 1998a;
see also Majewski et al.\ 1999d, Johnston et al.\ 1999a) is likely not
peculiar to the present epoch.

\section{Photometry}

The observations obtained for this project were accumulated over
several observing runs at the Las Campanas Observatory, when small
blocks of observing time were available (due to airmass, twilight, and
weather considerations) during other programs.  We include data from
both the Swope 1-m (C40) and du Pont 2.5-m (C100) telescopes.
Observations on the C40 were made with the same SITE\#1 CCD and
filters as employed in Paper I.  On the C40, this CCD gives 23.8
arcmin per side field-of-view.  Data were taken during grey or
bright time on the nights of UT 10 March 1999 and 28 April to 3 May
1999.  Data on the former run were photometric, while the CCD fields
observed during the latter run were not photometric.  In most cases,
CCD fields overlapped with neighbors so that color/magnitude
consistency could be checked.  For each C40 field, exposures of 120,
120, and 1200 seconds were taken in each of the Washington $M$, $T_2$ and $DDO
51$ filters.  All frames were reduced with the stand alone version of
DAOPHOT II (Stetson 1992) which produces point spread function-fitting
(PSF) photometry.  Figure 1 shows the distribution of all detected
stars in celestial coordinates; the lines show the boundaries of the
various CCD frames ({\it solid} lines show frames taken during
photometric conditions and {\it dashed} lines show frames take during
non-photometric conditions).  The density of detected stars at
different points in Figure 1 are a function of the relative
contribution of Carina, the relative proximity to the Galactic plane
(Carina is at a Galactic latitude of $b=-22^{\circ}$), the inclusion
of both C40 and (deeper) C100 data, and the degree of cloudiness
and seeing, which affects the limiting magnitudes of the C40 data.

\placefigure{Majewski.fig1.eps}

The PSF-fit magnitude measures were calibrated against Geisler (1990)
standards.  For the data taken during photometric conditions,
photometric transformation equations including airmass, color terms and
nightly zero-point terms were determined.  We followed the calibration
procedures described in Majewski et al.\ (1994), using a similar matrix
inversion algorithm (Harris, Fitzgerald \& Reed 1981).  The resultant
transformation equations were applied to all of the photometric frames.
A comparison of instrumental magnitudes in the CCD frames taken during
cloudy weather to the fully transformed magnitudes on the photometric
frames allowed derivation of frame-by-frame color and magnitude offset
terms for the former data; thus the CCD data taken during
non-photometric conditions were locked into the system of the calibrated
photometric magnitudes. Figure 1 shows the geometry of photometric and
bootstrapped fields.  Note that the photometric frames were located in
the center field and in a ring of fields separated from the center.
Non-photometric frames overlapping multiple photometric frames were
matched to all simultaneously.  When final calibrations of all frames
were achieved, a comparison of the derived magnitudes for stars in all
overlapping regions between CCD frames showed no major discrepancies:
the mean frame-to-frame offsets were typically of order 0.01 magnitudes.
Nevertheless, for our final catalogues we adopted the magnitude measures
for multiply photometered stars from the photometric frames over the
non-photometric frames, whenever possible.

The du Pont 2.5-m observations were made on the night of UT 11 March
1999 with the new Wide Field Camera (WFC) system built by Ray Weymann
and collaborators. The WFC delivers a useful circular field-of-view some
23 arcmin in diameter.  Four fields were observed: one centered on
the Carina core, two 50 arcmin along the major axis and outside the
tidal radius (adopting the structural parameters found by IH95), and one
along the minor axis at the same distance.  The locations of these
frames are shown by the circular fields in Figure 1.  During the early
operation of the WFC a residual misalignment of the field flattening
lens with respect to the instrument axis shifted the center of symmetry
of optimal focus slightly away from the center of the CCD frame, leaving
the edge of one quadrant with somewhat deteriorated image profiles of
sufficient severity that PSF-fitting photometry produced unsatisfactory
results, even after allowing for PSF variation with a quadratic
dependence on position in the CCD frame.  We decided to optimize our
DAOPHOT solutions to give good results over a large fraction of the CCD
field and sacrifice the ability to work with the bad portion of the
image, rather than trying to salvage the bad part of the field at the
expense of less than satisfactory PSF-fitting over the entire field.
Thus, as may be seen in Figure 1, the C100 data show a decline in the
number of detections to the upper left of each circular field.  While
this compromise means we lose stars from our survey, there should be no
preference for losing giants compared to dwarfs.

While the C100 data were taken during photometric conditions, limited
access meant there was no opportunity to obtain corresponding
calibration frames.  Fortunately, however, several of the C100 frames
overlap with the photometrically calibrated C40 grid, and from these
overlaps transformation equations could be derived for the C100 data.
The latter were applied to all C100 frames, whether they overlapped the
C40 data or not. Note that in the case where a star was photometered on
more than one set of CCD frames, a weighted average of the magnitudes
from the different frames was taken.

The photometric errors in the C40 and C100 data as a function of
magnitude for each filter are shown in Figure 2a for the C40 data and
Figure 2b for the C100 data. It can be seen that the C40 data, in
particular, show a wide range in quality.

\placefigure{Majewski.fig2.eps}

We remove from further consideration all detected objects with
non-stellar image profiles.  This was determined by deriving the running
mean (in a 50 star ``boxcar'' filter) of the DAOPHOT II $\chi$ and sharp
parameters as a function of magnitude and rejecting $3\sigma$ outliers
from this mean for the C40 data.  However, because of the problems with
the image quality on the C100 frames, we took a more conservative
rejection limit of $2.3\sigma$.  In addition, at this point we exclude
all stars that have magnitude errors in any filter that are larger than
0.1 magnitudes.  The latter cut is effectively one in magnitude (Figure
2) for each CCD field.

The $(M-T_2, M)_0$ color-magnitude diagram (CMD) for the C40 data for the
area shown in Figure 1 is shown in Figure 3a; the total CMD for the C100
data, which probe some 2 magnitudes deeper, is shown in Figure 3b.  Each
star has been corrected for reddening based on its celestial coordinates
and a comparison to the Schlegel et al.\ (1998) reddening maps.  The
precision of the photometry is typically about 0.04 magnitudes at $M =
19$ for the C40 data (but with a wide spread about this depending on the
particular frame -- see discussion of the relative magnitude limits
below), and 0.03 magnitudes at $M = 19$ for the C100 data.  The C100
data go deeper than the Carina horizontal branch (HB) at $M\sim20.5$,
while the Carina red clump is just barely detected in the C40 data.
Figures 3c and 3d show the CMDs for the stars actually used in our
survey , after application of the selection criteria described above.

\placefigure{Majewski.fig3.eps}

\section{Identification of Carina Giant Star Candidates}

Our strategy for identifying likely extratidal giant stars associated
with Carina tidal debris involves the application of two basic
criteria: (1) stars must have magnesium line/band strengths consistent
with those for giant stars with the abundance of Carina, and (2) stars
must have combinations of surface temperatures and apparent magnitude
consistent with the red giant branch of Carina.  We apply these
criteria in succession: 

\subsection{Giant Star Discrimination in the Two-Color Diagram}

Paper I describes the method by which dwarf/giant separation can be
achieved through the three filter imaging technique employed here.  The
basis of our technique lies in the sensitivity of the $DDO51$ filter to
the MgH band (bandhead at 5211 \AA\ ) and Mg b triplet near 5150 \AA\ 
(McClure 1976).  These magnesium features are sensitive to stellar
surface gravity (primarily) and temperature and abundance (secondarily)
in later type stars.  When combined with the wideband $M$ and $T_2$
filters of the Washington system, the $DDO 51$ filter is especially
useful for discriminating giant stars from foreground dwarfs on the
basis of differences in their respective $M-DDO51$ colors at a given
$M-T_2$ color.  The former color measures the strength of the magnesium
line/band strength (where $M$ acts as a suitable ``continuum'' measure
for comparison to $DDO 51$; Geisler 1984), while $M-T_2$ is sensitive
primarily to stellar surface temperature (and is almost a linear scaling
of $V-I$; Paper I).  With this photometric system, it is possible with
great efficiency to isolate giant stars at the distance of Carina with
small telescopes, even in bright, moonlit skies as we had for the Carina
observations here.

In Figure 4 we show the dereddened two-color diagram for both our C40
and C100 data, after pruning the sample with the error and image shape
criteria described above.  Figure 4 shows the characteristic
``elbow-shaped'' locus of dwarf stars (Paper I), which typically have
the largest magnesium absorption at any given temperature.  The region
enclosed by the box drawn with thick solid lines is the general area in
the two-color diagram inhabited by evolved, cool stars more metal-poor
than [Fe/H]$ \sim -0.5$ (Paper I).  The curved loci of giant stars of different
abundances, as determined from the synthetic photometry of Paltoglou \&
Bell (1994) and presented in Paper I are overlaid for comparison.  We
note that Carina is established to have [Fe/H]$ = -1.99$ with a
dispersion of 0.25 dex, based on spectroscopic observations of 52 giants by
Smecker-Hane et al.\  (1999).  The diagonal, blue boundary of the ``giant
region'' we have selected here is a somewhat conservative compromise to
produce relatively uncontaminated giant candidate samples, while not
sacrificing too many lower luminosity, bluer giants: The line is
approximately parallel to the center of the near-solar metallicity dwarf
locus, but offset by about +0.1 mag in ($M-DDO51$) to account for
typical magnitude errors at the faint end of the data sets.  We now
consider only stars in this delimited giant region as our first
selection for metal-poor giants in our survey fields.  As will become
evident below, this giant star ``bounding box'' selects not only Carina
stars on the RGB, but also Carina red clump stars.

\placefigure{Majewski.fig4.eps}

While the goal of our three filter photometry is to cast exclusively for
giant stars associated with Carina, our selection criterion in Figure 4
will also bring some contaminants into our net.  Obviously, we will
catch {\it any} giants with metallicities approximately like that of Carina.
There will also be some dwarf contamination due to several-$\sigma$
photometric errors scattering dwarfs into our giant star selection box.
Finally, as may be seen from the lines in Figure 4, metal-poor subdwarfs
with [Fe/H] $\lesssim -2.5$ also get pulled into our net.  We expect the
number of the latter type stars to be quite small, based on the very
small fraction of halo stars with metallicities this poor.  From Reid \&
Majewski (1993), the number of halo stars expected down to $V=20$ over
2.2 deg$^2$ is about 590. However, only a small fraction ($\lesssim
10\%$) of these halo stars will be cool enough (spectral type K and
later) to have an $(M-T_2)_0$ color which would place them in the giant
bounding box, and only about 8\% of {\it these} would be expected to
have metallicities as low as [Fe/H] $\lesssim 2.5$, according to Beers
(1999) and Norris (1999). This leaves an expected level of contamination of
$\lesssim 5$ metal-poor subdwarfs in our entire survey area. We conclude
that the vast majority of brighter non-Carina stars we select with only
a color-color criterion will be random field giants, while at faint
magnitudes we will pick up some dwarfs with extreme photometric errors.
Eventually, either of these two types of contaminants will be readily
identifiable through spectroscopy by their radial velocities (field
giants) or line strengths (photometric error dwarfs).  In another part
of our halo observational program, we have done a search for tidal
stellar debris from the Magellanic Clouds (Majewski et al.\ 1999a; see
Majewski et al.\ 1999c).  This part of our program includes both a
photometric search for giants, as we have done here, as well as
follow-up spectroscopy of the giant candidates.  It is worth pointing
out that in our Magellanic Cloud survey, which has identical photometric
material to that which we have used here, we have found a very high
success rate in the fraction of our giant candidates that we find to
have spectroscopic line strengths and velocities like metal poor, halo
giants.  We expect similar, or better, success rates here since in the
Magellanic Cloud work we used only aperture photometry, not PSF-fitting
photometry as we used here, and in that other work we also allowed a more
liberal selection (a larger ``giant box'') in the two-color diagram.

As a final check on the quality of the dwarf/giant discrimination with
our photometric technique, we consider the sample of 23 candidate Carina
RGB stars observed spectroscopically by Mateo et al.\ (1993).  They
chose these stars to lie near the top of the Carina RGB in the $(B-V,V)$
CMD, and found that 17 of the candidates had radial velocities
consistent with Carina membership, while six had heliocentric velocities
clustered around 0 km s$^{-1}$, consistent with their being foreground
dwarfs (see their Figures 3 and 4). All of these stars were photometered
by us, and we find that all 17 of the Carina RGB members (from Mateo et
al.) are clearly giants in our two-color diagram (Figure 5, filled
circles), while all 6 of the ``foreground'' stars of Mateo et al.\ are
clearly dwarfs (open circles). We note that even while our photometry
for several of the stars observed by Mateo et al.\ had errors that were
too large to keep them in our formal sample, the colors of these stars
still lie in the proper part of the color-color diagram, as seen in
Figure 5. This comparison gives confidence that with our technique we
can easily separate metal-poor giants from the typical foreground dwarf
to produce very ``clean'' lists of RGB candidate stars suitable for
followup spectroscopy.

\placefigure{Majewski.fig5.eps}

\subsection{The Color-Magnitude Locus of the Carina Red Giant Branch}

We now use our own photometry of Carina {\it itself} to establish the
expected location of associated evolved stars in the color-magnitude
diagram.  Figure 6 shows the color-magnitude distribution of all stars
selected as giants in Figure 4, but within 10 arcmin (roughly the core
radius) of the center of Carina for both the C40 and C100 data.  Note
that we measured the center of Carina from our own data by fitting
marginal distributions, but, as our determined center agreed to within 1
arcmin of the IH95 determination, for consistent comparisons we adopted
the IH95 Carina center for all calculations from here on.  We apply the
10 arcmin radial cut here to ensure that we obtain as pure a sample of
bound Carina stars as possible for defining the Carina RGB region in the
CMD.  As can be seen by Figure 6, the selection of ``giant star
candidates'' by the color-color technique seems to do a reasonable job
of isolating a relatively ``clean'' sample of evolved stars: Very few
stars fall outside the general region dominated by the Carina RGB in the
CMD.  Those stars falling away from the Carina RGB may either be dwarf
stars that failed our giant discrimination due to photometric error or
intrinsic properties ([Fe/H] $\lesssim -2.5$ dwarfs that show $M-DDO51$
colors of moderately metal-poor giants), they may be field giant stars,
or they may be Carina giants with several-$\sigma$ errors in their
photometry.

\notetoeditor{Please place Figures 6 and 7 on the same or facing pages
  if at all possible} 
\placefigure{Majewski.fig6.eps}

Based on the location of the primary locus of Carina RGB stars in Figure
6, we may now apply a second, {\it color-magnitude} selection criteria
to our giant star candidate sample, since, presumably, any RGB stars
associated with Carina, no matter how far from the core of Carina and
within our data set, should resemble RGB stars in the CMD of the Carina
core.  We note that the expected timescale ($\lesssim 1$ Gyr) for tidal
drift from Carina within the angles we survey are shorter than the
enrichment timescale (Gyrs).  Moreover, at present there is no evidence
for an age-metallicity relation among the variously aged populations in
Carina (Smecker-Hane et al.\ 1994, Da Costa 1994).  We also expect stars
in tidal tails to be lying within a few physical tidal radii (a few kpc)
of the Carina core along the line of sight; these differences in
distance would be virtually indistinguishable with our photometry.  Only
in very particular circumstances -- e.g., looking along Carina's orbit
-- would we expect to see tidal debris to be highly elongated along the
line of sight (for an example of the typical geometry of streamers
around a satellite see Johnston 1998, Figure 3).


From the Carina core RGB distributions in Figure 6 we define a {\it CMD
bounding box}, shown by the solid lines.  This region (assembled from
the combination of three second order polynomials) was defined to
contain the bulk of the Carina core RGB locus, as well as the apparent
red clump on the bluest end.

This same box may now be applied to the {\it entire} giant star
candidate sample over 2.2 deg$^2$ from \S 3.1 to pick from among
them those that, in addition to being pre-selected as evolved stars by
their colors, {\it are of the correct magnitude for their color} to be
associated with Carina (i.e., of an appropriate abundance/distance
combination).  Note that the actual size and shape of the bounding box
utilized here does not really matter as long as it is applied
consistently at all places in our survey mapping (Figure 1); in \S
3.4 we account for the level of contamination by background/foreground
stars, which scales with the size of the box.  A box that is too large
simply allows more contamination into our final selection of
Carina-associated stars.  While this translates into a lower efficiency
for follow-up spectroscopy of the selected sample, the increased
contamination level may be removed in a statistical way by appropriate
background subtraction.  On the other hand, a box that is too
restrictive means that more Carina-associated stars may be lost, and may
decrease the signal-to-noise of our extended stellar population
discussed below.  Again, we have attempted to compromise between these
extremes, except that we erred on the side of making the box a ``loose
fit'' to the RGB locus to account for the fact that extratidal stars may
acquire {\it slightly} different mean distances as they get drawn out with
different energies, ahead of or behind the parent object (e.g., as
suggested by the bridge/tail description of Toomre \& Toomre 1972).

\subsection{``Carina-Like'' Giant Candidates in the Two-Color and Color-Magnitude Diagrams}

Figure 7 shows the primary locations of the core Carina RGB, as selected
by the CMD-selection box in Figure 6.  We see that the color-color
distribution of these stars is smaller than the entire ``giant box'' in
the two-color diagram, and one could conceive of narrowing the
color-color selection criterion further by collapsing the giant box
around the Carina locus.  We do not do so here, however.

\placefigure{Majewski.fig7.eps}
 
We now apply the selection criteria defined by the bounding boxes in
both Figures 4 and 6 to the entire sample of stars in the C40 and C100
data sets to define the sample of most likely Carina-associated stars.
In Figure 8 we show the CMDs of all C40 and C100 stars satisfying the
two-color selection criterion in Figure 4 -- i.e., all stars selected as
evolved stars of similar metallicity to Carina in our entire survey
area.  Even when all evolved stars from the entire survey are included
in the CMD, the dominant CMD structure, particularly for cool stars, is
the Carina RGB.  We include in Figure 8 the CMD selection criterion
given by the bounding box selected in Figure 6. We note that it encloses
most of the apparent Carina RGB, as expected. However, a notable
exception is the trio of stars at $M_o\sim17.5$ and $(M-T_2)_o>1.85$:
Though these three stars appear to be an extension of the Carina RGB
bounding box, the latter does not extend red enough to include them
because there were no examples of such relatively rare stars in the
Carina core. Though we have formally excluded these stars in our
analysis, we do consider these three stars as likely Carina-associated
stars that we would have found with an appropriately-extended CMD
bounding box. We return to discuss these stars in our spectroscopic
tests in \S 3.5.

\placefigure{Majewski.fig8.eps}

\subsection{Sky Distributions and Evaluation of Giant Background Level}

Figure 9a shows the distribution on the sky of all stars selected by our
combined color-color-magnitude selection.  The central fall off in the
concentration of Carina giant candidates is obvious, but the fall off
does not truncate completely, and, indeed, the density of candidates
seems to flatten out and extend not only beyond the core radius, but the
tidal radius as well.  By comparison, we show in Figure 9b the stars
that have similar color-color characteristics as Carina giants (i.e.,
within the box in Figure 4) but {\it outside} the CMD bounding box in
Figure 6, i.e., stars that would be metal-poor giants, but generally at
different distances than Carina.  The latter show no central
concentration, but rather, for the most part, the (expected) random,
flat distribution of halo field giants.

\placefigure{Majewski.fig9.eps}

At first, the similarity of the distribution of the relative stellar
density in the outer parts of Figure 9a and Figure 9b may appear to be a
cause for concern.  Some of the similarity is related to the differing
depths of the individual CCD fields, which modulate both the number of
detected Carina-associates and non-associates. Moreover, there could be
some ``spill-over'' of true Carina stars to just outside of our
selection criteria, which moves them from panel 9a to panel 9b.  But of
most concern to our purpose here is what fraction of the extratidal
stars in Figure 9a are likely to be real and how many are expected to be
``interlopers'' -- e.g., (1) dwarf stars that are accidentally selected
to be ``Carina-like'' giants due either to photometric errors or
extremely low [Fe/H], or (2) actual giant stars that happen to have the
correct color/abundance/distance characteristics that place them into
our sample?  We must evaluate the expected level of contamination from
interlopers, and we do so by monitoring the ``background level'' of such
stars as a function of magnitude.

Before continuing, however, we must stress that the sky distributions of
candidate giant stars as illustrated in Figure 9 are modulated by the
variable depth of our data across the entire survey area and Figure 9a,
even if accurately depicting the existence of extratidal Carina debris,
cannot be interpreted as a mapping of the true relative density of that
debris.  Our analysis must proceed by taking into account the relative
depths of our somewhat inhomogeneous data set.  We do so by analyzing
the survey with four different magnitude limits.  At each magnitude
limit, we include only those survey areas that are complete to that
depth.  The net effect of this approach is that with an increasing
magnitude limit we cover less area on the sky, but we are able to
recover greater densities of potential Carina-associated stars in the
smaller areas because we probe further down the RGB.  The goal of
analyzing different magnitude-limited data sets in this way is a fair
appraisal of not only the expected contamination levels, but the true
relative sky densities of giant stars, while taking maximal advantage of
the area covered at various depths.

Figure 10 shows the sky distributions of color-color-magnitude selected
Carina-associated giant candidates taking into account the magnitude
limits of the frames.  For comparison, Figure 11 shows the same for all
stars selected as metal-poor giants in the color-color diagram, but
which are not along the Carina RGB in the color-magnitude diagram (i.e.,
presumably metal-poor giants at different distances than Carina).  In
Figure 10a and perhaps Figure 10b, it can be seen that the brighter,
candidate Carina-associated stars do seem to show an overall radial
drop-off from the core, but one that continues beyond the nominal tidal
radius of IH95 (compare to the presumably ``random field'' star
distributions in Figures 11a and 11b).  For simplicity during the
remainder of the discussion in this Section, we will refer to these
stars outside the IH95 tidal radius as ``extratidal'', though we
acknowledge the controversy regarding the true tidal radii of dwarf
galaxies like Carina, as discussed in \S 1.  Unfortunately, it is more
difficult to follow any apparent radial trend in the deeper survey
fields shown in Figures 10c and 10d, because of the poor radial sampling
in the placement of the fields.  On the other hand, it can be seen that
the total number of extratidal giant candidates in Figures 10c and 10d
outnumber the stars in the same regions of the sky in Figures 11c and
11d, respectively.  This is significant because the relative areas in
the color-magnitude diagram from which the extratidal giants are culled
is much smaller in Figure 10 than in Figure 11.  Thus, it would appear,
we are seeing a significant excess in extratidally-positioned stars at
just the colors and magnitudes expected for Carina-associated
populations.

\notetoeditor{Please place Figures 10 and 11 on the same or facing pages
  if at all possible} 
\placefigure{Majewski.fig10.eps}
\placefigure{Majewski.fig11.eps}

To put the latter assessment on a more quantitative footing, we assess
in Figures 12 and 13 the background contribution of field giant stars
(and the expected small contribution of foreground dwarfs from
photometric error and extreme subdwarfs) to our counts of candidate
Carina-associated giants.  The foundation of our background analysis is
the assumption that the distribution of random halo field giants should
be relatively smooth and slowly varying with distance.  Indeed, if the
Galactic halo follows anything close to an $R^{-3}$ power law, as is
widely assumed (and reported from the most recent surveys of blue
horizontal branch stars; see a recent summary in Sluis \& Arnold 1998),
the counts of halo giants per unit solid angle should be flat, modulo
second order effects relating to possible metallicity gradients (which
are generally not found in the outer halo; Searle \& Zinn 1978, Zinn
1985, Carney et al.\ 1990, Armandroff, Da Costa \& Zinn 1992, Rich
1998).  In our case, we count giant stars {\it already pre-selected (on
  the basis of their position in the color-color diagram) to be
  metal-poor}; if we adopt a counting filter in the CMD with a shape
matching the CMD selection box in Figure 6 (which follows the outline of
an [Fe/H]$\sim -2$ RGB) our magnitude counts of these giants translate
more or less directly into counts by distance modulus.

\notetoeditor{Please place Figures 12 and 13 on the same or facing pages
  if at all possible} 
\placefigure{Majewski.fig12.eps}
\placefigure{Majewski.fig13.eps}

Thus, we offset the CMD bounding box of Figure 6 by 0.5 magnitude
intervals, and at each offset position count the number of giants
satisfying the color-color criterion shown in Figure 4.  These counts
are summarized for each of our four magnitude-limited data sets in
Figure 12 (which shows all metal-poor, color-color selected giants) and
Figure 13 (which shows only those metal-poor, color-color selected
giants outside the IH95 tidal radius).  Note that the actual filter used
for each panel in Figures 12 and 13 was modified to take into account
the varying depths of the four magnitude-limited data sets.  For
example, in the $M<19.3$ data set, the bottom of the CMD bounding box is
truncated precisely at $M=19.3$.  In turn, the $M < 19.8$ data set is
analyzed with the appropriately truncated CMD bounding box at $M=19.8$,
and etc.  For each magnitude-limited data set, the modified, truncated
bounding box is the one offset and used to produce the giant count
histograms in Figures 12 and 13.  Note that only offsets in the
direction of brighter magnitudes make sense, as offsets in the fainter
direction incorrectly evaluate the numbers of stars due to sample
incompleteness at the faint end.  The maximum negative magnitude offset
was given by the bright-end, CCD-saturation limit of the survey.

The main feature to note in each panel of Figure 12 and 13 is the
relatively flat contributions of stars at magnitudes brighter than the
Carina RGB.  Indeed, under the assumption that the majority of these
stars are giant stars and not dwarfs with large photometric errors or
low [Fe/H], {\it the high degree of flatness in the histograms strongly
  supports an $R^{-3}$ distribution for Galactic halo field giants (or at
  least metal-poor giants).}  We assume this flatness
persists through the magnitude range dominated by the Carina RGB
($\Delta M = 0$) in our survey, and adopt the mean level of the flat
distribution as our background level of field halo giants and other
interlopers in our ``Carina-associated'' giant sample at $\Delta M = 0$.

As the magnitude offsets approach 0 in each case shown in Figures 12 and
13, we see a sudden rise in the numbers of giants counted in the
shifting counting box.  The peak histogram values centered on $\Delta M
= 0$ give the total number of stars in the Carina CMD bounding box as
originally centered on the Carina RGB.  But the sharpness of the rise
seems to vary among the different samples. This is because there is some
overlap of the shifted box with the true Carina RGB for small magnitude
offsets and the maximum vertical extent of the bounding box varies:
$\Delta M = 1.3$, 1.7, 2.1 and 2.5 magnitudes for the $M < 19.3$, $M <
19.8$, $M < 20.3$ and $M <20.8$, respectively.  If the
``Carina-contaminated'' bins less than these $\Delta M$ are ignored, we
may determine the mean expected background contribution to our candidate
Carina-associated stars from the various test offsets of the CMD
bounding box in Figures 12 and 13.  These data are included in Table 1.

\placetable{Table 1}

Figure 12 shows that when all color-color selected metal-poor giants
are considered regardless of their sky position in our survey, the
number of expected contaminants lying within the Carina RGB is rather
small: $<4\%$ for all four magnitude limits explored.  This
suggests that, unless some peculiar problem is affecting our candidates
specifically at the color-magnitude location of the Carina RGB, we might
expect relatively high Carina membership probabilities from a
spectroscopic follow-up study of these candidates (an expectation that 
is supported by our successful dwarf/giant discrimination of the Mateo
et al.\ 1993, stars discussed at the end of \S 3.1 and shown in
Figure 5, and by our spectroscopy in \S 3.5 below).  Moreover, from the data
in Figure 13 and Table 1 we see that the excess of candidate extratidal
Carina-associated giants is at the level of 3.7$\sigma$ or more for each
of the four magnitude-limited data sets we explore.

A final observation to be noted from Figure 10 is that the extratidal
distribution of Carina-like giant candidates appears rather isotropic,
however, our field coverage is not ideal for assessing this.  In contrast,
Kuhn et al.\ (1996) report no extratidal Carina extension perpendicular
to its major axis, but note that they explore only two minor axis fields
2$^{\circ}$ away from the Carina center. We also note some
interesting similarities in the angular distribution of our
Carina-associated giant candidates, particularly those in the $M < 19.8$
and $M < 20.3$ data sets (Figures 10b and 10c), and the isopleths
published by IH95.  In particular, the various spurs of higher density
extending off of the IH95 central Carina contours and extending past the
tidal radius (especially the spur to the southwest, but also the several
other spurs at other position angles) are rather similar to such
features in our data.  Perhaps this is not surprising, as IH95's
starcount analysis was essentially limited to counting Carina RGB
members, albeit about a magnitude deeper than we have here.
Nevertheless, the apparent general agreement in the two rather different
analyses is encouraging.

\subsection{Spectroscopy}

Table 1 and Figures 10 and 13 suggest that we have found a significant
``extratidal'' population around Carina.  A radial velocity survey to
confirm whether these stars are indeed Carina-associated and, if so, to
understand their velocity characteristics, is an obvious next
observational step.  While we have been unable to do such a study, we
have managed to secure spectra of two of our Carina giant candidates --
both exterior to the IH95 tidal radius -- during twilight time on nights
allocated for other programs on the C100. With the remaining available
telescope time, we also decided to observe two of the brighter,
$(M-T_2)_o>1.85$ stars that lie outside of the Carina CMD bounding box,
but that do appear to be at the very tip of the Carina RGB (see
discussion in \S 3.3). These stars yielded much better S/N spectra in
the observing time available than the other two stars almost a magnitude
fainter.  One of these brighter stars is inside the tidal radius, while
the other lies exterior.  All four spectra were taken on the nights of
UT 27 Aug to 1 Sep 1999 with the Modular Spectrograph. The wavelength
range spans from approximately H$\beta$ to H$\alpha$ at $\sim$ 1 \AA\ 
pixel$^{-1}$.  The spectrographic set-up and the radial velocity
reduction procedures have been described elsewhere (Majewski et al.\ 
1999d). We present the results of this analysis, and other data about
the stars including the positions, the angular separation from the
center of Carina, the magnitude and color, our derived radial velocities
and the height of the radial velocity cross-correlation peak (see
Majewski et al.\ 1999d), in Table 2.  We show the positions of the four
stars observed spectroscopically as triangle symbols in Figure 10a.

\placetable{Table 2}

From repeat measures of previously well-observed stars during this
observing run, we have determined our external, random and systemic
velocity errors on the Carina candidate spectra to be $10-15$ km
s$^{-1}$; this is sufficient to check association with Carina, but not
good enough to make conclusions regarding possible differential velocity
structure.  The heliocentric radial velocity of Carina is 224$\pm3$ km
s$^{-1}$ (Mateo 1998) with a spread in the velocities of individual
carbon stars and giants of about $\pm15$ km s$^{-1}$ (Lynden-Bell,
Cannon \& Godwin 1983; Mateo et al.\ 1993).  Star C1407251, a bright
giant candidate located within 16 arcmin of the Carina core, has a
velocity that agrees with the Carina velocity and is certainly a member.
Star C2103156, the bright giant star candidate that lies outside the
IH95 tidal radius, gives a spectrum that looks remarkably metal-poor and
very similar to that of star C1407251; the radial velocity for star
C2103156 lies within 2$\sigma$ of Carina's systemic velocity and we
consider this giant candidate a likely member of the Carina system.  Our
spectra of stars C2103156 and C2501583, the two extratidal stars that
were observed and that lie inside the Carina CMD bounding box, are also
rather devoid of significant absorption lines which again suggests an
association with Carina. Unfortunately, however, the combination of no
strong lines and a weak and noisy signal make it hard to get a good
radial velocity for these stars: We obtain marginal cross-correlation
peaks that we generally regard as unacceptably small ($<0.3$) and
indicative of a several times larger random velocity error.  Nevertheless,
the derived heliocentric velocities are very close to Carina's
(with C2501927 almost an exact match) and give rather poor matches
to the expected velocity of foreground dwarfs, $\sim 20$ km s$^{-1}$
(Mateo et al.\ 1993). We conclude the that latter two stars are far more
likely to be associated with Carina than to be foreground stars.

We regard this small, but important test of four of our identified
Carina-associated RGB candidates as vindication of our approach, and as
support for our claims that the distribution of candidate
Carina-associated giants in Figures 9 and 10 is most likely to reflect a
real extended structure of the Carina dwarf.  We hope to make further
observations of additional extratidal candidate stars with a larger
aperture telescope.

\subsection{Radial Profiles}

We present ``Carina RGB'' starcounts as a function of elliptical annuli,
along with the sampled areas in each annulus, in Table 3.  The shape of
the elliptical annuli were adopted from the parameters given by IH95,
namely an ellipticity of 0.33 at a position angle of 65$^{\circ}$.  We
space the width of our annular counting bins at one-fifth the major axis
tidal radius given by IH95, except we use two times finer resolution in
our first four bins.  The $r_{inner}$ and $r_{outer}$ listed in Table 3
correspond to the inner and outer radius of the annuli along the major
axis.

\placetable{Table 3}

We convert the annular counts to densities (taking into account the
actual survey area covered within each annulus) subtract the mean
density of the background counts (derived from data in Tables 1 and 3
and presented in Table 3), and present the resultant radial densities
(per arcmin$^{2}$) in Figure 14.  To improve our signal-to-noise we
combine into two bins our outermost eight annuli in Figures 14 and 15.
The difference in the relative numbers of stars at each radius for our
different magnitude-limited samples merely reflects the increase in the
relative density of stars as a function of survey depth.  In general,
the counts from the different magnitude-limited data sets track each
other at all radii (but, of course, the four data sets are not
completely independent), though the $M < 19.3$ and $M < 19.8$ show more
Poissonian scatter at large radii.  We also include in Figure 14 the
Carina count data as presented by IH95 (their Table 3), which has a
magnitude limit almost another magnitude deeper than our $M < 20.8$
sample and which shows a commensurately higher density scaling.  Our
counts roughly track the IH95 counts, as well as the similarly deep Kuhn
et al.\ (1996) starcount data also presented in Figure 14.  Note that
the IH95 data looses signal-to-noise after about 40 arcmin, at which
point the background correction becomes critical, and IH95 limited their
presentation to this radius (so we do so here as well).  Even before 40
arcmin, however, the IH95 data show the effects of decreased
signal-to-noise. On the other hand, our $M < 20.8$ data have reasonable
signal-to-noise to almost 80 arcmin, at which point we are limited only
by the extent of our survey sky coverage.  Thus, our technique could
potentially probe the extended structure of Carina to even greater radii
than we have done here.

\notetoeditor{Please place Figures 14 and 15 on the same or facing pages
  if at all possible} 
\placefigure{Majewski.fig14.eps}
\placefigure{Majewski.fig15.eps}

In order to compare results more readily, we normalize the relative
densities in our four data sets to the IH95 data (Figure 15a).  We
normalize near the radius corresponding to our third annulus where our
data have the highest counts (signal-to-noise). For the Kuhn et al.\ 
(1996) data, we normalize to the IH95 counts at the Carina core.  The
various data sets show general agreement in the character of the radial
profile, especially within 20 arcmin.  Moreover, as found by IH95 and
hinted at in the counts by Demers et al.\ (1983), our data show a break
in the fall-off rate of the radial counts near 20 arcmin, and this
slower rate of decline continues to the radial limit of our survey area.
However, the level of the counts in the IH95 data tend to be several
times higher than ours in the outer radii of overlap (from about 13 to
40 arcmin), though this is where the IH95 data show large uncertainties
and are most affected by their adopted background levels.  Within our
own survey there is a trend in that the data sets with brighter
magnitude limits have faster fall-offs at large radii than do the deeper
data sets.  This is likely a result of the fact that our brighter data
sets face the problem of quantization noise at smaller radii than do the
fainter data sets.  Thus, for purposes of analysis at the largest radii,
we take the $M<20.3$ and $M<20.8$ data sets as most likely to represent
the true density profile (Figure 15b), though even these have some
quantization noise at the outermost extent of our survey area.

The King profile (King 1962, 1966) fit by IH95 to the central region of
their Carina data is also shown in Figure 15a, and highlights the
dramatic break in our radial density rate of decline after about 20
arcmin.  Our beyond-the-break counts of giants also approximately match
the Kuhn et al.\ (1996) starcount data (which monitor excess populations
only beyond the Carina break radius), though the latter also show more
apparent scatter at large radii.  This general match to the Kuhn et al.\ 
data to the limit of our survey is in spite of the fact that the Kuhn et
al.\ data correspond only to fields along the Carina major axis.  One
might interpret this to suggest that our adoption of uniformly shaped
and oriented elliptical annuli has little effect on the derived radial
profiles, but we note that in our deepest data sets the sampling of
azimuthal angles around Carina narrows to a range dominated by the
Carina major axis, similar to the sampling by Kuhn et al.

\subsection{Mass Loss Rate}

Models of the radial distribution of the stars around a tidally
disrupting satellite, e.g., by Johnston et al.\ (1999b, 2000), show
characteristics very similar to those shown in Figure 15 (compare to
Figure 15 of Johnston et al.\ 1999b).  The Johnston et al.\ model
demonstrates that the break point results from the contribution and
eventual dominance of unbound stars.  Dashed lines are included in
Figure 15 past the tidal radius to represent various $r^{-\gamma}$-laws
discussed by Johnston et al.\ (1999b, 2000).  We present in Figure 15b
only our deeper samples, for clarity in the comparison.  It can be seen
that our deeper survey data have a radial fall-off somewhere between
$\gamma =1$ and $2$ to the limit ($\gtrsim 80$ arcmin) of our areal
coverage.

Given the match of our data as presented in Figure 15 to the model
predictions of Johnston et al.\ (1999b), we proceed for now under the
assumption that past the radial profile break we are seeing unbound,
extratidal debris, and calculate the mass loss rate using the formalism
outlined in Johnston et al.\ (1999b).  Note that even with "perfect"
observations of their simulations Johnston et al.\ (1999b, 2000) found
that they could only recover the rate of destruction of their satellites
to within a factor of two. Hence the dominant source of uncertainty in
our own calculation will be from the inherent simplicity of the model
rather than observational errors, and the number we derive should be
taken as an order-of-magnitude estimate of the destruction rate rather
than a definitive measurement.

Because the Johnston et al.\ (1999b) formalism assumes complete area
sampling in the derivation of the relative numbers of stars within
certain annuli, we scale our counts of stars at each of our annuli by
the ratio of the total elliptical area to the amount of that annulus we
actually surveyed.  Under the Johnston et al.\ nomenclature, we adopt
$r_{break}$ as occurring between our sixth and seventh annuli (23
arcmin), the radius to which the extratidal debris is well-defined as
$R_{xt}=64$ arcmin (between our thirteenth and fourteenth annuli), and
take for Carina's orbital parameters $g(\theta)=1$ and $T_{orb} = 2\pi
R_{GC}/(200 \rm{km/s})$ with $R_{GC}=101$ kpc; this yields a mass-loss
rate of $\begin{pmatrix} {df \over dt}\end{pmatrix}_1 = 0.27$
Gyr$^{-1}$.  This rather high value is relatively insensitive to the
actual outer radial limit we take for the extratidal population,
$R_{xt}$.  We note that Johnston et al.\, who used the surface
brightness at the location of $r_{break}$ in the IH95 data to estimate
the mass-loss rate with an alternative computational method, obtained
$\begin{pmatrix} {df \over dt}\end{pmatrix}_2 < 0.33$ Gyr$^{-1}$, an
upper limit in agreement with our value.  Carina and Ursa Minor have the
largest estimated upper limits for mass loss rates among the dwarf
galaxies discussed by Johnston et al.\ (all Milky Way dSph's excluding
Sagittarius).  The implication of a mass-loss rate of this order of
magnitude is that Carina will not likely survive another Hubble time.
Extrapolating backwards in time, one comes to the conclusion that Carina
has likely lost a significant amount of mass already, and might be
expected to sport significant tidal tails (see \S 4.3).

\section{Discussion}

\subsection{Summary of Results}

Our goal was to find evidence for, and begin mapping, tidal debris in
the Carina dwarf galaxy by way of a search for Carina-associated giant
stars to well beyond the nominal Carina tidal radius.  We have used two
criteria to select stars that are candidate giant stars associated with
the Carina dwarf spheroidal: (1) colors in the ($M-T_2$, $M-DDO51$)
plane, where we are able to isolate evolved stars with [Fe/H] of order
that of Carina, and (2) positions in the CMD that are similar to those
of Carina giant branch and red clump stars.  We check the background
level of halo field giants, metal-poor subdwarfs masquerading as giants,
and other possible interlopers, and find that they make a minor
contribution to our signal.  The latter stars, which we expect mainly to
be random halo field giants, show a flat magnitude count slope, which
suggests that they follow an $R^{-3}$ law, as is commonly adopted for
the halo.

We derive the radial profile of candidate Carina-associated giants, and
find a break in the counts at about 20 arcmin, near the tidal radius
derived by IH95 (who have better sampling and signal-to-noise in the
Carina core).  A well-established, $3.7\sigma$ excess of
Carina-associated giant candidates is found beyond this radius, and
spectroscopy of several of these stars verifies that they represent a
real extended structure of Carina.  The beyond-the-break stars show an
$r^{-\gamma}$ decline in their radial fall-off, with $1<\gamma<2$, that
is of a form similar to the predictions for unbound stars in tidally
disrupting systems (Johnston et al.\ 1999b).  Our excess of giant stars
outside the King model cut-off radius may well represent stars that have
recently been stripped from Carina due to Galactic tidal forces.  Or
they may represent still bound, retrograde revolving counterparts to
those stars that {\it were} tidally stripped due to the fact that they
happened to be in prograde orbits when they became extratidal (Innanen
\& Papp 1979).  Alternatively, if, as discussed in \S 1, modest amounts
of dark matter prohibits the production of tidal tails, our Carina
radial profile to $> 80$ arcmin heralds the need for an explanation of
multiple component structures in dwarf spheroidals like Carina.  Each of
these alternatives has a distinct, kinematical signature that would be
recognizable with an appropriate radial velocity survey.

Note that our search technique identifies {\it actual} dwarf
galaxy-associated giant candidates and thus we are able not only to find
radially-averaged galaxy profiles, but to make two-dimensional maps of the
local overdensity of extended debris.  In the case of Carina, since we
observe viable candidates to the edges of our survey area, several times
beyond the King model cut-off radius, it is possible that we are only
seeing the beginnings of a wider diaspora of Carina stars that, if tidal
debris, will eventually sort by relative energies into classic tidal
tails at larger distances from Carina.  This hypothesis must be
followed up both by casting a wider photometric net for extratidal
Carina giants at greater angular separations from the Carina core, and
by testing whether the dynamics of present and future samples of
``extratidal Carina-associated giants'' have the proper, rotation-like
radial velocity signature predicted for tidal debris.  Unfortunately, we
have been able to do the proper spectroscopic confirmations for only
three extratidal stars, and so it is beyond the capabilities of the
present paper to settle the thorny and weighty issues concerning dark
matter, tidal debris and the true location of dwarf spheroidal tidal
radii outlined in the Introduction.  Instead, we choose to end our
discussion by raising additional intrigue as to how Carina's extended
populations may affect study of two of its neighbors -- the Magellanic
Clouds and the Milky Way.

\subsection{Possible Connection to Apparent Tidal Debris Near the Magellanic Clouds}

In a separate contribution from our program to study substructure in the
halo, we discuss a targeted search for tidal stellar debris from the
Magellanic Clouds (Majewski et al.\ 1999a; see Majewski et al.\ 1999c).
The latter work includes observations in a partially-filled ring of
fields encircling both Clouds.  We have found coherent radial velocity
structures among the distant giants identified in almost every field we
have surveyed, which strongly suggests that there is tidal debris widely
dispersed across the stretch of sky we have sampled in that survey
(i.e., an envelope from $250^\circ$ to $320^\circ$ in Galactic longitude
and from $-18^\circ$ to $-55^\circ$ in Galactic latitude).  However, the
strongest signal we have encountered -- both in the excess in the
density of giants identified as well as in the tightness of the
coherence of the radial velocities of these stars -- is among a set of
six fields spanning a 15$^{\circ}$ arc located $\sim18^{\circ}$ from the
center, and to the northeast, of the LMC (we term these fields LMC-NE
here).  The LMC-NE fields are placed directly between the LMC and
Carina, with the surveyed arc of fields slicing across the arc
connecting the LMC and Carina, 8/10 of the way from the LMC to Carina.
It is too soon to ascribe the coherently moving stars in the LMC-NE
fields as Magellanic in origin; however, their spatial and velocity
distribution are not inconsistent with model expectations (Majewski et
al.\ 1999a) for Magellanic debris, with a possible additional
contribution of a moving group of stars from the LMC polar ring
described by Kunkel et al.\ (1997).  These findings may be relevant to our
findings here since the LMC-NE fields showing the distant moving group
stars get as close as $3^\circ$ from the center of Carina.

Note also that Carina lies near the Magellanic plane, along with the
Clouds, Ursa Minor, Draco and a number of globular clusters, and some or
all of these objects have been proposed to represent chunks of debris
from the break-up of a formerly larger progenitor Magellanic system
(Lynden-Bell 1976, Kunkel 1979, Palma, Majewski \& Johnston 1999).  In
this scenario, these dwarf galaxy/globular cluster chunks would likely
be awash in a debris stream of stars also pulled out of the progenitor.
Thus, if Carina itself has an origin {\it as} tidal debris, we might
expect coherent groups of stars nearby, whether drawn from Carina
directly, or not.  An argument against a picture as just painted is that
tidal dwarf galaxies are not expected to contain dark matter (Barnes \&
Hernquist 1992, Moore 1996, Burkert 1997, Klessen \& Kroupa 1998),
whereas large dark matter contents ratios have been used to explain the
high velocity dispersion of Carina stars (e.g., Mateo et al.\ 1993).

Whether Magellanic in origin or not, there {\it is} a blanket of
coherently moving stars in the outer halo in this general direction of
the sky and detected within 3$^{\circ}$ of Carina.  It is worth checking
whether this blanket extends to the position of Carina and contributes
to the extratidal giant candidates we have found around Carina. However,
given that there is a radial fall-off of stars with distance from {\it
  Carina}, we might not expect {\it all} of the extratidal Carina giants
to be contributed by the LMC-NE feature.  Clearly a more extensive
survey of these stars from the LMC to Carina is needed.

\subsection{Implications for the Structure and Origin of the Milky Way Halo}

If Carina is losing stars, then the Milky Way halo is gaining them.
Because of its age distribution, Carina presents an interesting case for
the accretion of stars in the halo.  If disintegrating, Carina should
{\it presently} be contributing predominantly intermediate age ($\sim 7$ Gyr)
stars (Mould \& Aaronson 1983, Mighell 1990) with a small admixture of
stars from its old (12-15 Gyr) and young (2-3 Gyr) burst populations
(Smecker-Hane et al.\ 1994, Grebel 1998, Mateo 1998).  The derived
proportional integrated star formation for these populations, based 
on Carina's present ratios of different aged stars, varies
among authors but averages to ratios of old:intermediate:young
approximately as 0.2:1.0:0.1.  The youngest stars accreted from Carina
may be comparable to the Preston et al.\ (1994) blue metal-poor stars,
of which about half are thought to be relatively young stars from
accretion events (Preston 1999, personal communication).

A comparison of the numbers of such young stars in Carina to the
number in the Galactic halo has been used to provide an upper limit
on the contribution of stars to the
Milky Way by Carina or Carina-like, accreted galaxies:  
Unavane, Wyse \& Gilmore (1996) calculate that at most
approximately 60 dwarfs with the mass and metallicity of Carina could
have been accreted by the Galactic halo, and would now account for
a total of $\sim3\%$ of the mass of the halo.  However, such a calculation
assuming a {\it static} Carina may greatly underestimate the 
potential contribution of matter to the Galactic halo via the
accretion of dwarf galaxies.

The present mass loss rate we have determined (\S 3.7) suggests that Carina
is now losing of order 27\% of its mass every Gyr, and, since that rate
was determined from the current distribution of luminous matter under a
scenario where light traces mass, that fractional mass loss rate may be
adopted for both the dark and luminous matter.\footnote{The luminosity of 
Carina is 4.3 x 10$^5$ L$_{\odot}$ and its total estimated mass is 1.3 x 10$^7$
M$_{\odot}$, which yields an integrated $M/L$ of 31 (Mateo 1998).  }
If we assume this
fractional mass loss rate as typical over the life of Carina, then we
approximate the mass of Carina $N$ Gyr ago to be $(0.73)^{-N}$ larger
than at present.  Thus, we find that Carina was approximately a factor
of 2, 10, and 100 times larger at the times of the bursts occurring
approximately 2, 7, and 14 Gyr ago.  Thus, Carina's predominant stellar
contribution to 
the Milky Way may have been in the form of {\it old} stars from its
first starburst. 

This mass loss rate also suggests that, if having been maintained for the
past Hubble time, accretion of Carina {\it alone} would have contributed 
about 6\% of the Galactic halo's mass, and Carina itself
would now be reduced to 1\% of its original
mass.  Interestingly, Klessen \& Kroupa (1998) find in N-body
simulations of the tidal interaction of a satellite with a massive
galaxy that the models converge to a dwarf remnant that has 1\% of the
mass of the initial satellite.

If Carina has been losing mass at this prodigious rate, then it has lost
nearly all of the stellar component formed more than 7 Gyr ago. The loss
of so much of the old stellar population could dramatically distort the
observed star formation history (SFH) with respect to the actual SFH:
The fact that 80\% of the stars currently in Carina appear to be younger
than the burst of star formation that occurred 7 Gyr ago (Hurley-Keller
et al.\ 1998) could be ascribed to the fact that 90\% of the mass of
proto-Carina had already been accreted by the Milky Way by that time.

\acknowledgments

We thank Andi Burkert for helpful conversations.  Chris Palma is
gratefully acknowledged for his creation of the SKAWDPHOT program used
for the photometric transformation solutions.

We are especially thankful for the support given this project by the
Carnegie Observatories, both in Pasadena and at Las Campanas.  The
Director of the Carnegie Observatories, Dr. Augustus Oemler, has been
especially kind with the granting of discretionary time, and with
supporting SRM as a Visiting Associate of the Observatories.  This
telescope access has been particularly helpful as we have been
unsuccessful obtaining telescope time for this project elsewhere.

JCO and SRM acknowledge several grants in aid of undergraduate research
from the Dean of the College of Arts \& Sciences at the University of
Virginia.  SRM acknowledges partial support from an National Science
Foundation CAREER Award grant, AST-9702521, a fellowship from the David
and Lucile Packard Foundation, and a Cottrell Scholarship from The
Research Corporation.

\clearpage

\begin{center}
{\large\bf Captions}
\end{center}

\figcaption[Majewski.fig1.eps]{Map of all detected objects in our survey
  region centered on Carina.  Squared regions indicate boundaries of
  individual C40 frames, whereas round regions are those surveyed with
  the C100 WFC.  The ellipse is the 29 arcmin tidal radius at position
  angle 65$^{\circ}$ derived by IH95. Solid lined boundaries indicate
  frames that are fully photometric and tied to standard stars. These
  were used to bootstrap the photometry for the other fields shown with
  dashed line boundaries.\label{Majewski.fig1.eps}}

\figcaption[Majewski.fig2.eps]{Photometric errors for stellar objects in
  the survey fields as a function of magnitude for the (a) C40 and (b)
  C100 data.\label{Majewski.fig2.eps}}

\figcaption[Majewski.fig3.eps]{Dereddened $(M-T_2, M)_o$ color magnitude
  diagrams for the (a) C40 data and (b) C100 data.  In (c) and (d) only
  objects detected with stellar profiles and with magnitude errors less
  than 0.1 mag in all filters are included for the (c) C40 and (d) C100
  data.\label{Majewski.fig3.eps}}

\figcaption[Majewski.fig4.eps]{$(M-T_2, M-DDO51)_o$ diagrams for all of
  the (a) C40 and (b) C100 data in the present survey. Only objects
  detected with stellar profiles and with magnitude errors less than 0.1
  mag in all filters are included.  The {\it solid line} shows the
  bounding region we have employed to select metal poor giant star
  candidates.  The curves (adapted from Paltoglou \& Bell 1994) show the
  expected location of dwarfs and giants; for clarity we break up the
  curves by panel as (a) giants (with abundances, from top to bottom:
  [Fe/H]$=-3.0, -2.0, -1.0, 0.0$) and (b) dwarfs (with abundances from
  top to bottom: [Fe/H]$=-3.0, -2.0, -1.0, +0.5$) (see Paper
  I).\label{Majewski.fig4.eps}}

\figcaption[Majewski.fig5.eps]{(a) $(M-T_2, M)_o$ diagram for the C40
  CCD frame centered on Carina with the candidate Carina RGB stars from
  Mateo et al.\ (1993) marked.  The stars that have radial velocities
  measured by Mateo et al.\ consistent with Carina membership are
  indicated by filled circles, while the stars with radial velocities
  consistent with their being foreground dwarfs are indicated by open
  circles.  (b) $(M-T_2, M-DDO51)_o$ diagram for the same data. The
  spectroscopically confirmed Carina RGB stars lie comfortably within
  the giant box indicated in Figure 4, while the Mateo et al.\ 
  foreground dwarfs land precisely on the expected locus for
  $-1.0\le\rm{[Fe/H]}<+0.5$ dwarfs (see Figure
  4b).\label{Majewski.fig5.eps}}

\figcaption[Majewski.fig6.eps]{$(M-T_2, M)_o$ color-magnitude diagram
  for stars selected as metal-poor giants in Figure 4 and within 10
  arcminutes of the center of Carina.  Panel (a) shows the C40 data and
  panel (b) shows the C100 data.  The ``box'' shown by the solid lines
  is our CMD selection criterion for ``Carina-associated RGB
  stars''.\label{Majewski.fig6.eps}}

\figcaption[Majewski.fig7.eps]{$(M-T_2, M-DDO51)$ diagram with stars
  selected by the bounding box for the Carina RGB in Figure 6 shown as
  X's.  Panel (a) shows the C40 data and (b) shows the C100
  data.\label{Majewski.fig7.eps}}

\figcaption[Majewski.fig8.eps]{$(M-T_2, M)_o$ color-magnitude diagram
  for all stars selected by the two-color selection criterion shown in
  Figure 4.  The CMD bounding box defined from Carina stars within its
  core radius is shown by the solid lines.\label{Majewski.fig8.eps}}

\figcaption[Majewski.fig9.eps]{(a) Distribution on the sky of all stars
  selected as ``Carina-like'' giants by {\it both} the final color-color
  and color-magnitude selection criteria.  (b) Distribution on the sky
  of all stars with positions in the color-color diagram within the
  bounding box in Figure 4, but {\it not} within the color-magnitude
  bounding box defined in Figure 6.  As before, the {\it solid lines}
  indicate frames that were taken under photometric conditions and {\it
    dashed lines} indicate those that were not.\label{Majewski.fig9.eps}}

\figcaption[Majewski.fig10.eps]{Sky distribution of all stars selected
  as ``Carina-like'' giants by {\it both} the final color-color and
  color-magnitude selection criteria for the magnitude limits (a) $M <
  19.3$, (b) $M < 19.8$, (c) $M < 20.3$ and (d) $M < 20.8$.  In each
  panel we show the boundaries of, and candidates from, only those CCD
  survey areas that are complete to the magnitude limit indicated. The
  four triangles in panel (a) mark the positions of the four
  spectroscopically observed stars listed in Table 2 and discussed in \S
  3.5.\label{Majewski.fig10.eps} }

\figcaption[Majewski.fig11.eps]{Same as Figure 10, but for those stars
  selected as ``Carina-like'' giants by color-color, but {\it not}
  color-magnitude criteria.\label{Majewski.fig11.eps}}

\figcaption[Majewski.fig12.eps]{Counts of giant candidates selected as
  low-metallicity giants by the color-color selection criterion in
  Figure 4 as a function of the $M$ band magnitude offset, $\Delta M$,
  of the CMD bounding box in Figure 5.\label{Majewski.fig12.eps}}

\figcaption[Majewski.fig13.eps]{Same as Figure 12, but for extratidal
  giant candidates only.\label{Majewski.fig13.eps}}

\figcaption[Majewski.fig14.eps]{Radial profile of the density (per
  arcmin$^{2}$) of candidate Carina-associated giants for the $M<19.3$
  ({\it open triangles}), $M<19.8$ ({\it solid triangles}), $M<20.3$
  ({\it open circles}) and $M<20.8$ ({\it solid circles}) samples. For
  some points at small radii, the error bars are smaller than the
  plotted symbol. The {\it asterisks} are the background-subtracted
  counts as presented in IH95.  The {\it stars} show the results from
  Kuhn et al.\ (1996).  In the IH95 and our own data, we have combined
  some of the outer radial bins to increase
  signal-to-noise.\label{Majewski.fig14.eps}}

\figcaption[Majewski.fig15.eps]{(a) Radial profile of the density of
  candidate Carina-associated giants combined with IH95 starcounts and
  all normalized at 8.7 arcminutes.  Symbols are as in Figure 14.  The
  solid line shows the King profile as obtained from the data in IH95.
  The dashed lines show $r^{-\gamma}$ fall-offs in the ``extratidal
  domain'', with $\gamma =$1,2 and 3 and with normalization to our
  seventh bin.  We have combined some of the larger radius bins in our
  survey and in IH95 to increase signal-to-noise.  (b) Same as (a), but
  for only our $M<20.3$ and $M<20.8$ samples.\label{Majewski.fig15.eps}}

\clearpage

\begin{deluxetable}{ccccc}
\tablecolumns{5}
\scriptsize
\tablecaption{Total Carina-Associated Giant Candidates and Expected Background Counts\label{Table 1}}
\tablehead{
\colhead{Magnitude limit}  & \colhead{Total Counts} &  
\colhead{Background}  & \colhead{Extratidal Counts} & 
\colhead{Extratidal Background} }  
\startdata
$M < 19.3$       &    223       &  $7.8\pm2.9$ &     26            &  $7.1\pm2.8$ \\
$M < 19.8$       &    385       & $15.0\pm4.2$ &     53            & $12.7\pm3.1$ \\
$M < 20.3$       &    552       & $22.0\pm4.3$ &     89            & $18.0\pm3.9$ \\
$M < 20.8$       &    800       & $31.0\pm4.9$ &    117            & $26.2\pm4.7$ \\
\enddata
\end{deluxetable}

\begin{deluxetable}{rrrrrrrrr}
\tablecolumns{5}
\scriptsize
\tablecaption{Radial Velocities of Carina-associated Giant Candidates\label{Table 2}}
\tablehead{
\colhead{Name}  & \colhead{R.A.} &  
\colhead{Dec.}  & \colhead{$r$} &\colhead{$M_o$} & \colhead{$(M-T_2)_o$} & \colhead{$(M-DDO51)_o$} &
\colhead{R.V.}  & \colhead{X-corr} \\
\colhead{} & \multicolumn{2}{c}{(2000.0)} & \colhead{($\arcmin$)} &\colhead{} &\colhead{} & \colhead{} & \colhead{(km~s${}^{-1}$)} & \colhead{peak}}
\startdata
C2501583 & 6 38 22.6 & -51 10 59 &34& 18.32 & 1.68 & $+0.02$ & 287.4 & 0.24 \\
C2501927 & 6 38 37.0 & -51 16 23 &34& 18.50 & 1.58 & $+0.04$ & 223.1 & 0.20 \\
C2103156 & 6 45 31.0 & -50 49 26 &37& 17.76 & 2.00 & $-0.03$ & 250.7 & 0.61 \\
C1407251 & 6 40 08.8 & -50 57 11 &16& 17.62 & 1.95 & $-0.01$ & 233.1 & 0.77 \\
\enddata
\end{deluxetable}

\begin{deluxetable}{rrr@{~~~~~}lr@{~~~~~}lr@{~~~~~}lr@{~~~~~}l}
\tablecolumns{10}
\scriptsize
\tablecaption{Radial Counts of Carina-Associated Giant Candidates\label{Table 3}}
\tablehead{
\colhead{$r_{inner}$} & \colhead{$r_{outer}$}& 
\multicolumn{2}{c}{$M<19.3$} & \multicolumn{2}{c}{$M<19.8$} & 
\multicolumn{2}{c}{$M<20.3$} & \multicolumn{2}{c}{$M<20.8$}\\
\colhead{(arcmin)}  &  \colhead{(arcmin)}  &\colhead{count} & 
\colhead{arcmin$^2$} & \colhead{count} & \colhead{arcmin$^2$} & 
\colhead{count} & \colhead{arcmin$^2$} & \colhead{count} & 
\colhead{arcmin$^2$}}  
\startdata
  0.000     &  2.898     & 10 & \phn18  &   16 & \phn18  &  32 & \phn18  &   \phn54 & \phn18  \\
  2.898     &  5.796     & 37 & \phn53  &   68 & \phn53  &  97 & \phn53  &  159 & \phn53  \\
  5.796     &  8.694     & 58 & \phn88  &   84 & \phn88  & 124 & \phn88  &  190 & \phn88  \\
  8.694     & 11.592     & 41 & 124   &   68 & 124   &  91 & 124   &  145 & 124   \\
 11.592     & 17.388     & 35 & 354   &   60 & 354   &  82 & 333   &  107  & 234   \\
 17.388     & 23.184     &  \phn9 & 468   &   15 & 468   &  15 & 261   &   \phn13 &  \phn90   \\
 23.184     & 28.980     &  \phn8 & 639   &   22 & 639   &  23 & 423   &   \phn16 & 180   \\
 28.980     & 34.776     &  \phn8 & 693   &   15 & 693   &  17 & 531   &   \phn17 & 297   \\
 34.776     & 40.572     &  \phn6 & 801   &   10 & 738   &  18 & 558   &   \phn20 & 378   \\
 40.572     & 46.368     &  \phn2 & 855   &    \phn7 & 621   &  13 & 522   &   \phn20 & 423   \\
 46.368     & 52.164     &  \phn2 & 819   &    \phn2 & 621   &   \phn4 & 504   &    \phn\phn6 & 396   \\
 52.164     & 57.960     &  \phn3 & 891   &    \phn4 & 720   &  14 & 585   &   \phn26 & 477   \\
 57.960     & 63.756     &  \phn1 & 720   &    \phn1 & 612   &   \phn5 & 549   &    \phn\phn6 & 477   \\
 63.756     & 69.552     &  \phn1 & 450   &    \phn3 & 441   &   \phn6 & 288   &    \phn\phn6 & 288   \\
 69.552     & 75.348     &  \phn3 & 324   &    \phn9 & 324   &   \phn9 & 252   &   \phn11 & 252   \\
 75.348     & 81.144     &  \phn0 & 315   &    \phn2 & 315   &   \phn3 & 180   &    \phn\phn3 & 180   \\
 81.144     & 86.940     &  \phn0 & 126   &    \phn0 & 126   &   \phn0 & 117   &    \phn\phn1 & 117   \\
 86.940     & 92.736     &  \phn0 &  \phn81   &    \phn0 &  \phn81   &   \phn0 &  \phn81   &    1 &  \phn81   \\
 92.736     & 98.532     &  \phn0 &  \phn18   &    \phn0 &  \phn18   &   \phn0 &  \phn18   &    0 &  \phn18   \\
 98.532     & 104.328    &  \phn0 &   \phn\phn0   &    \phn0 &   \phn\phn0   &   \phn\phn0 &   \phn\phn0   &    \phn\phn0 &   \phn\phn0   \\
\hline
\multicolumn{2}{c}{total area (deg$^2$)}  & \multicolumn{2}{c}{2.18} & \multicolumn{2}{c}{1.96} & \multicolumn{2}{c}{1.52} & \multicolumn{2}{c}{1.16}  \\
\multicolumn{2}{c}{background (deg$^{-2}$)} & \multicolumn{2}{c}{3.58}& \multicolumn{2}{c}{7.65} & \multicolumn{2}{c}{14.47} & \multicolumn{2}{c}{26.72}  \\
\enddata
\end{deluxetable}

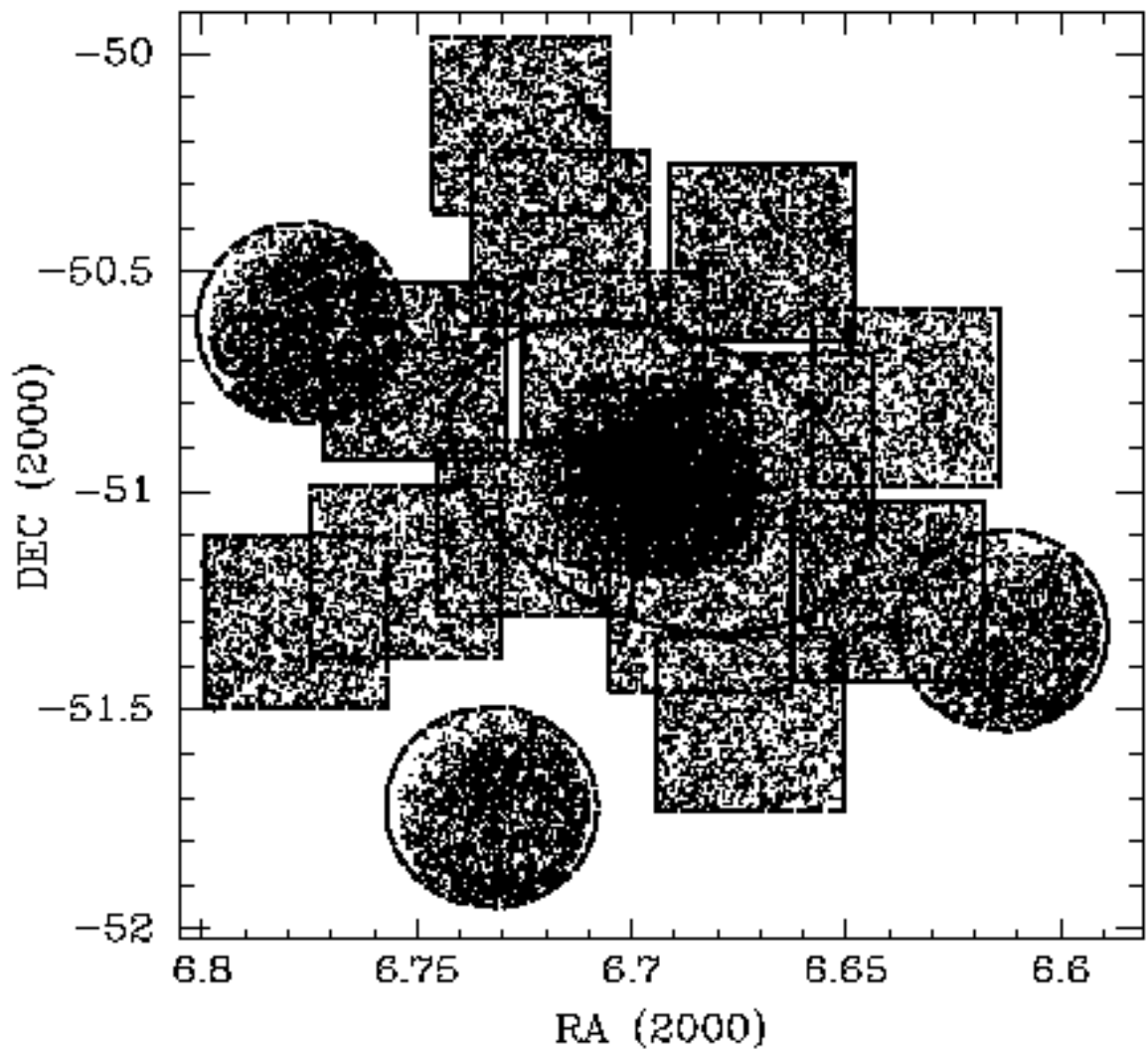
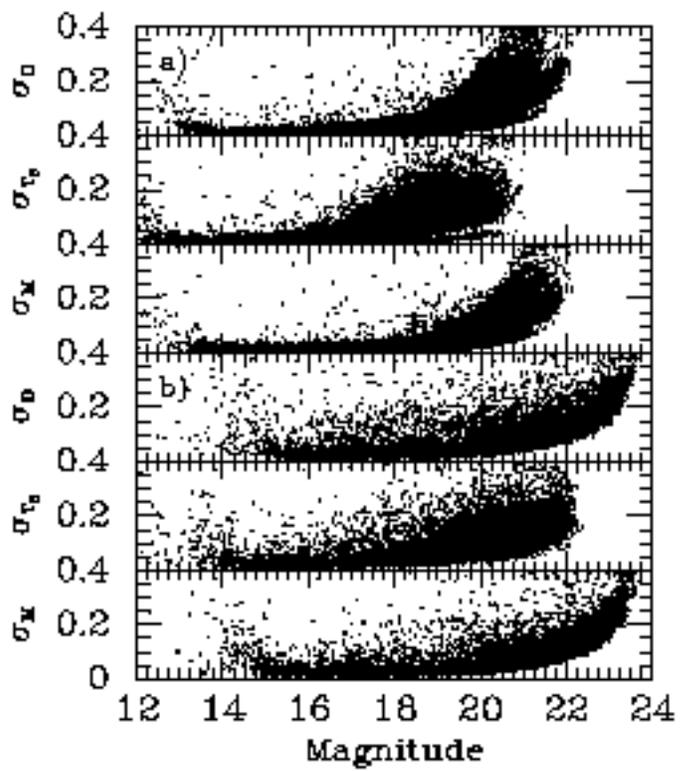
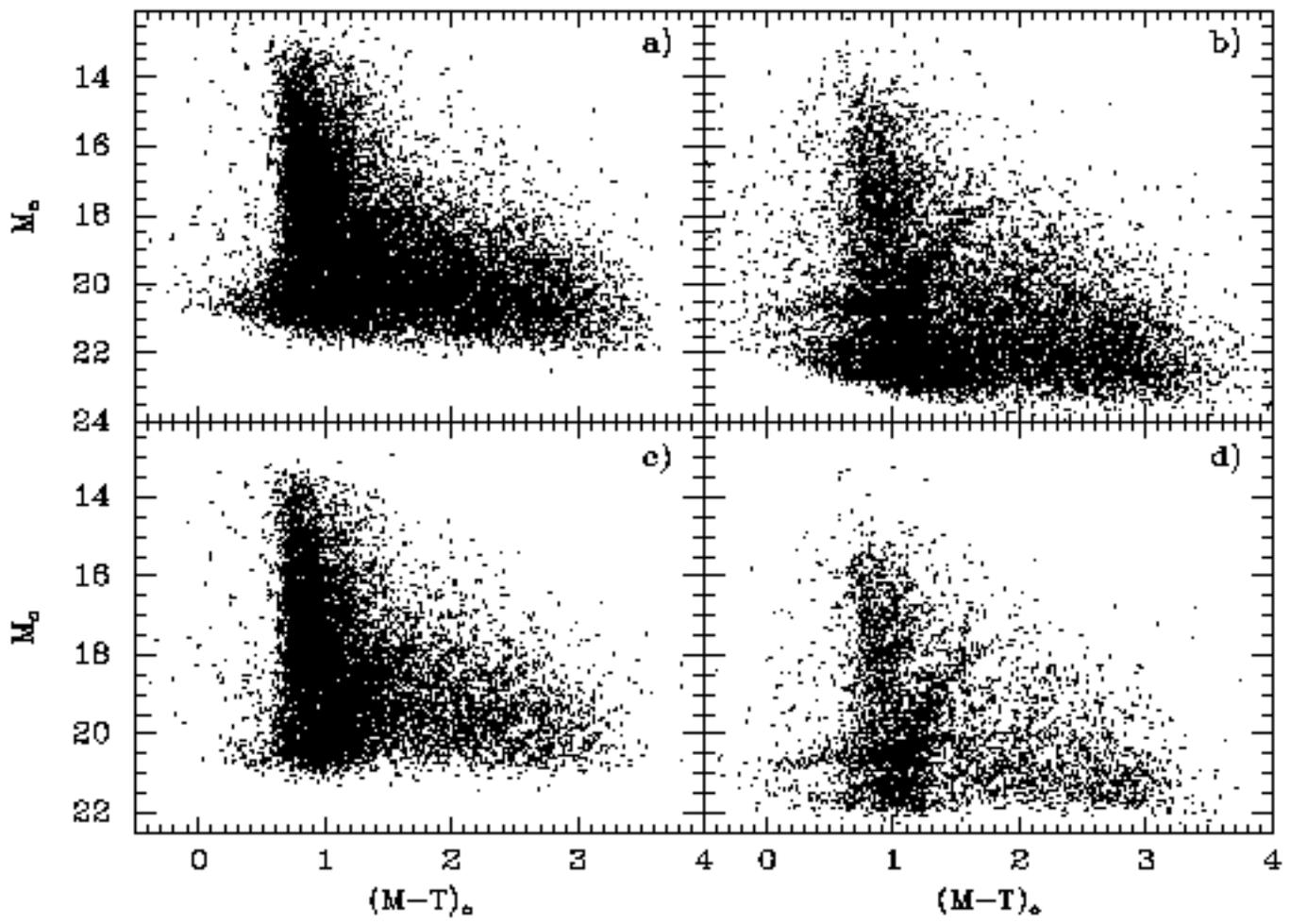
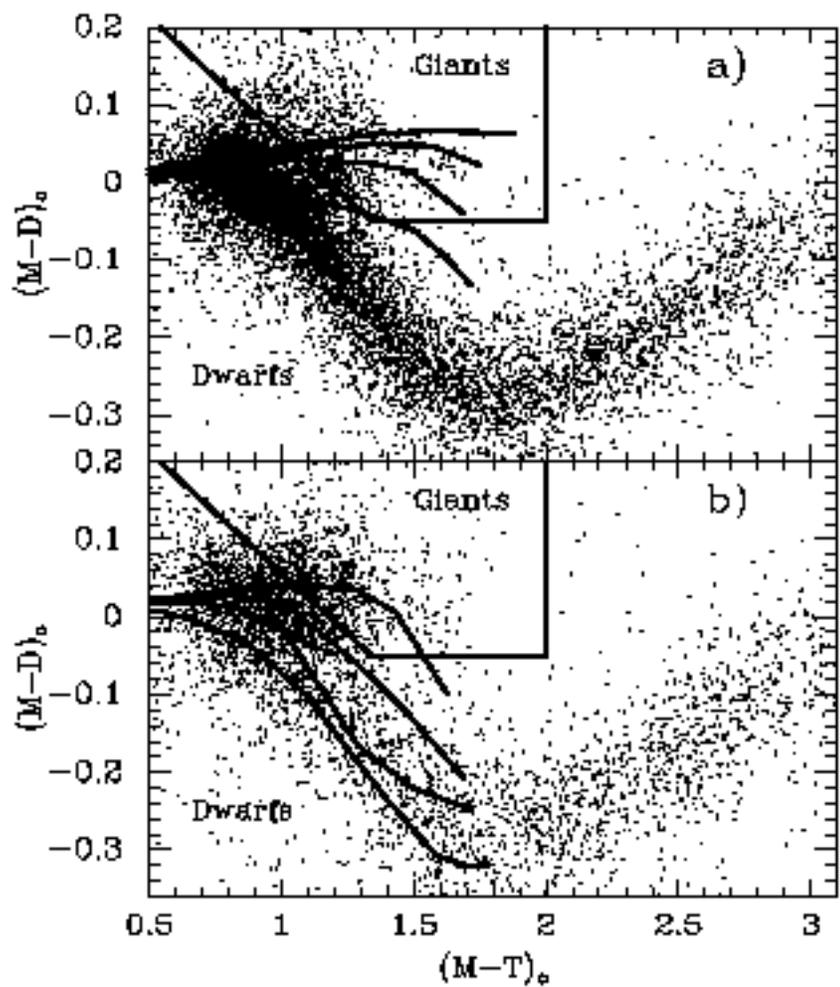
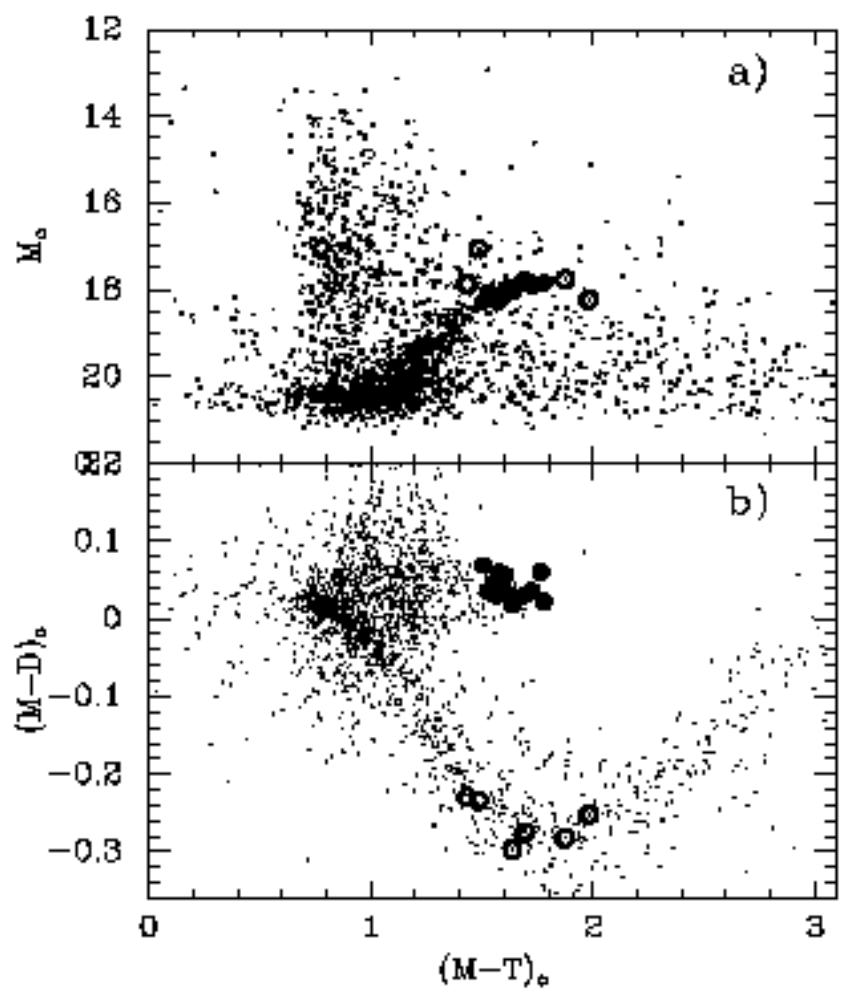
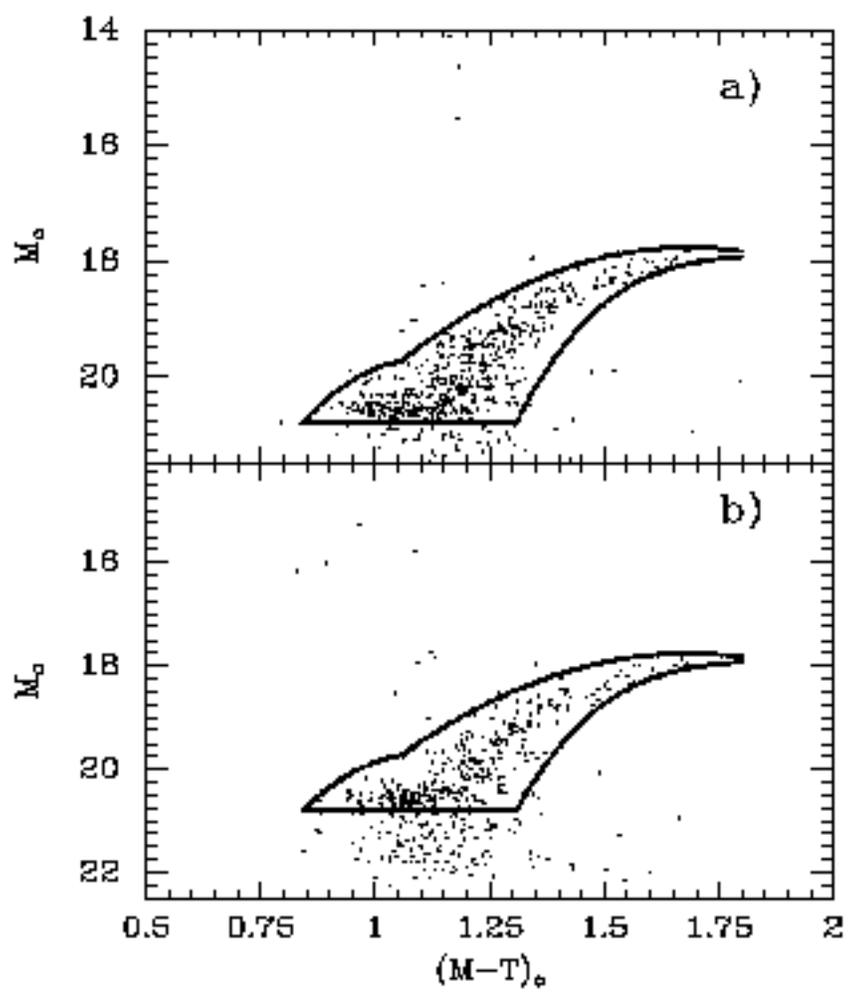
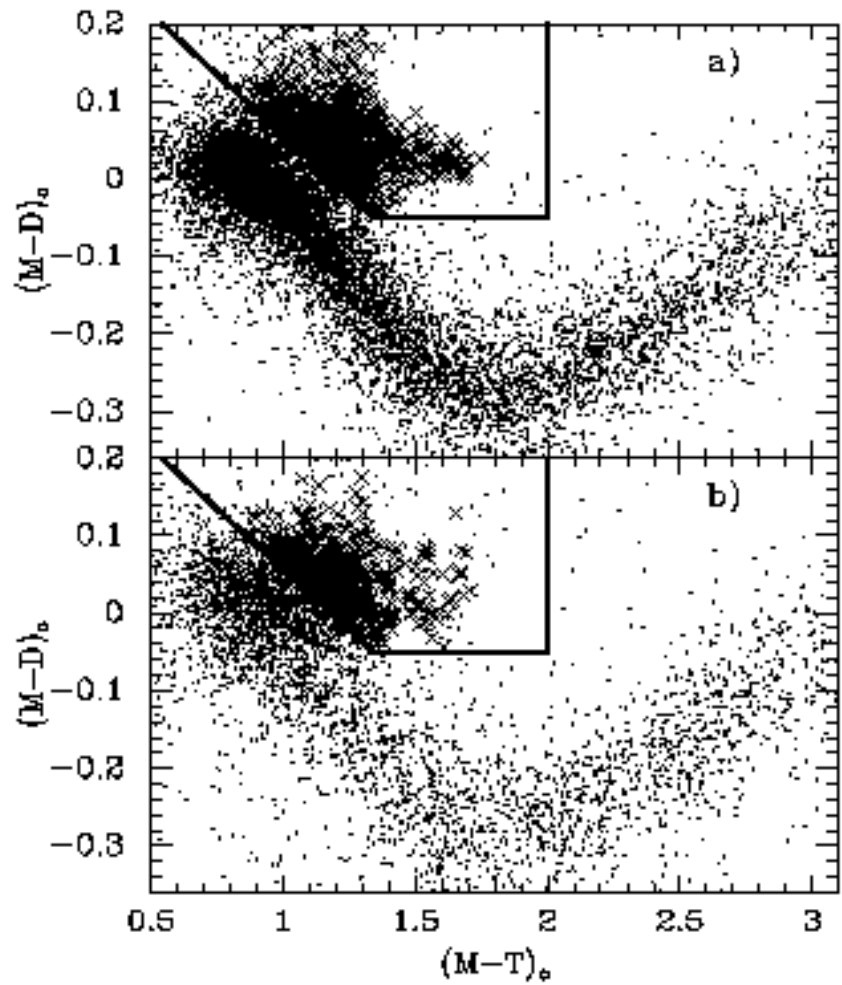
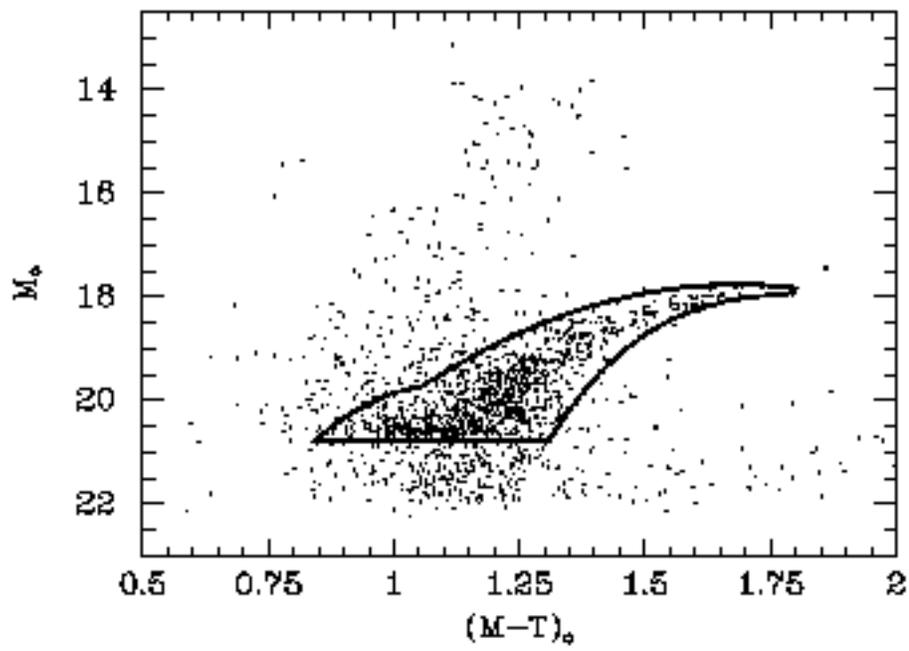
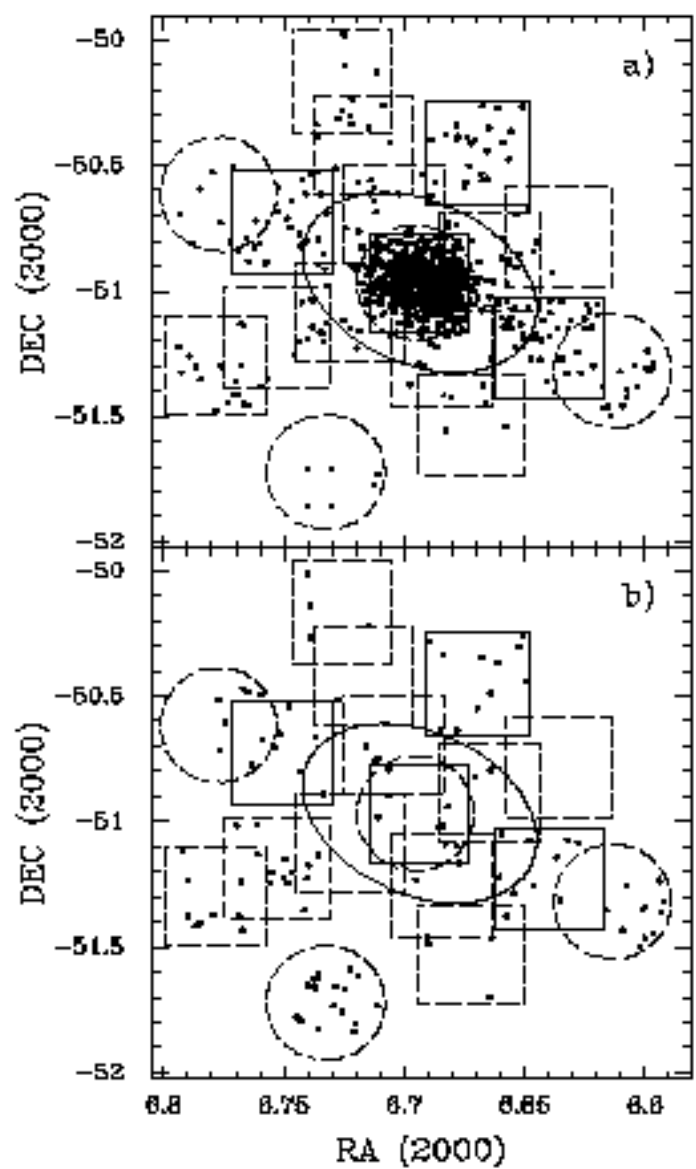
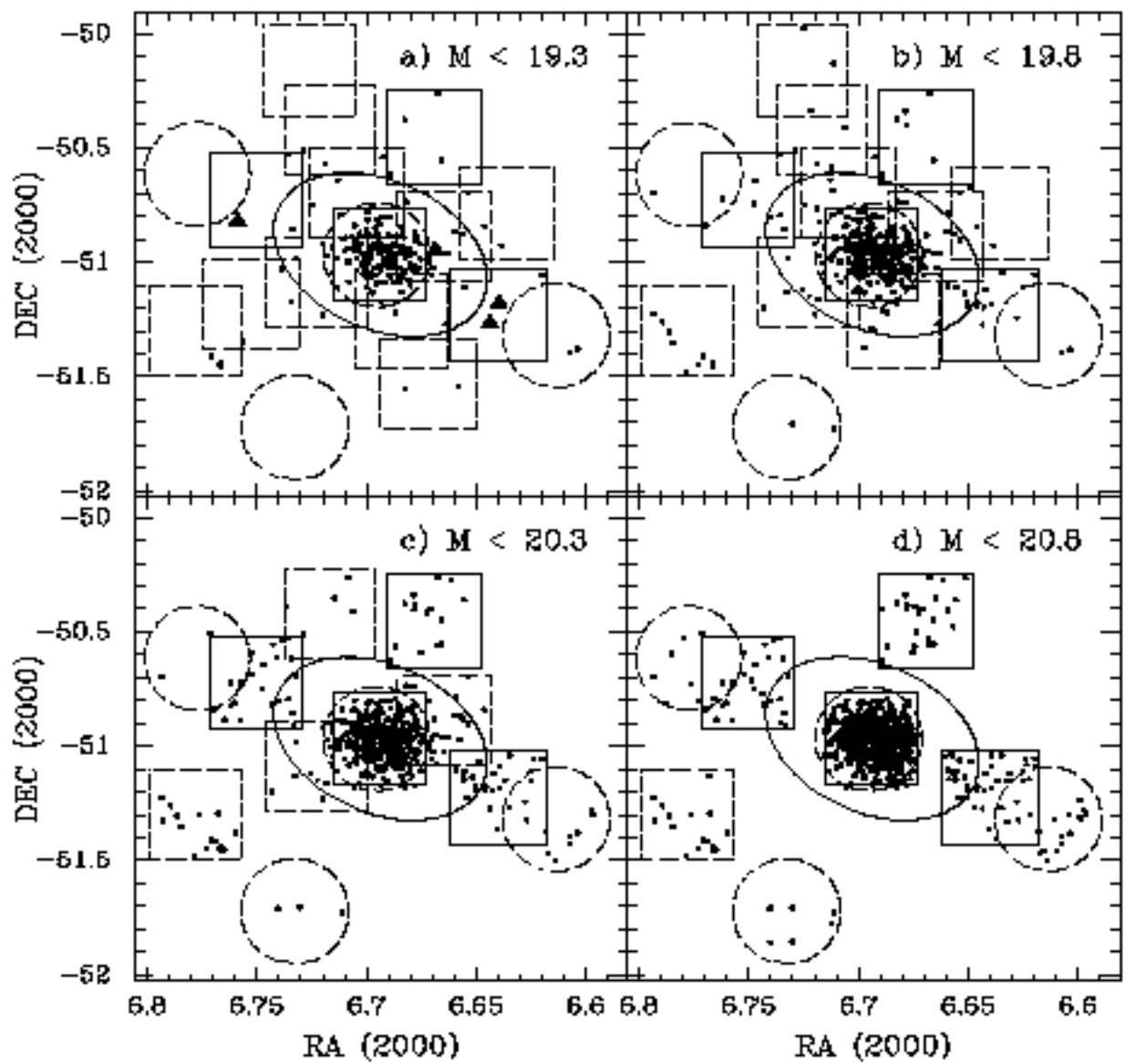
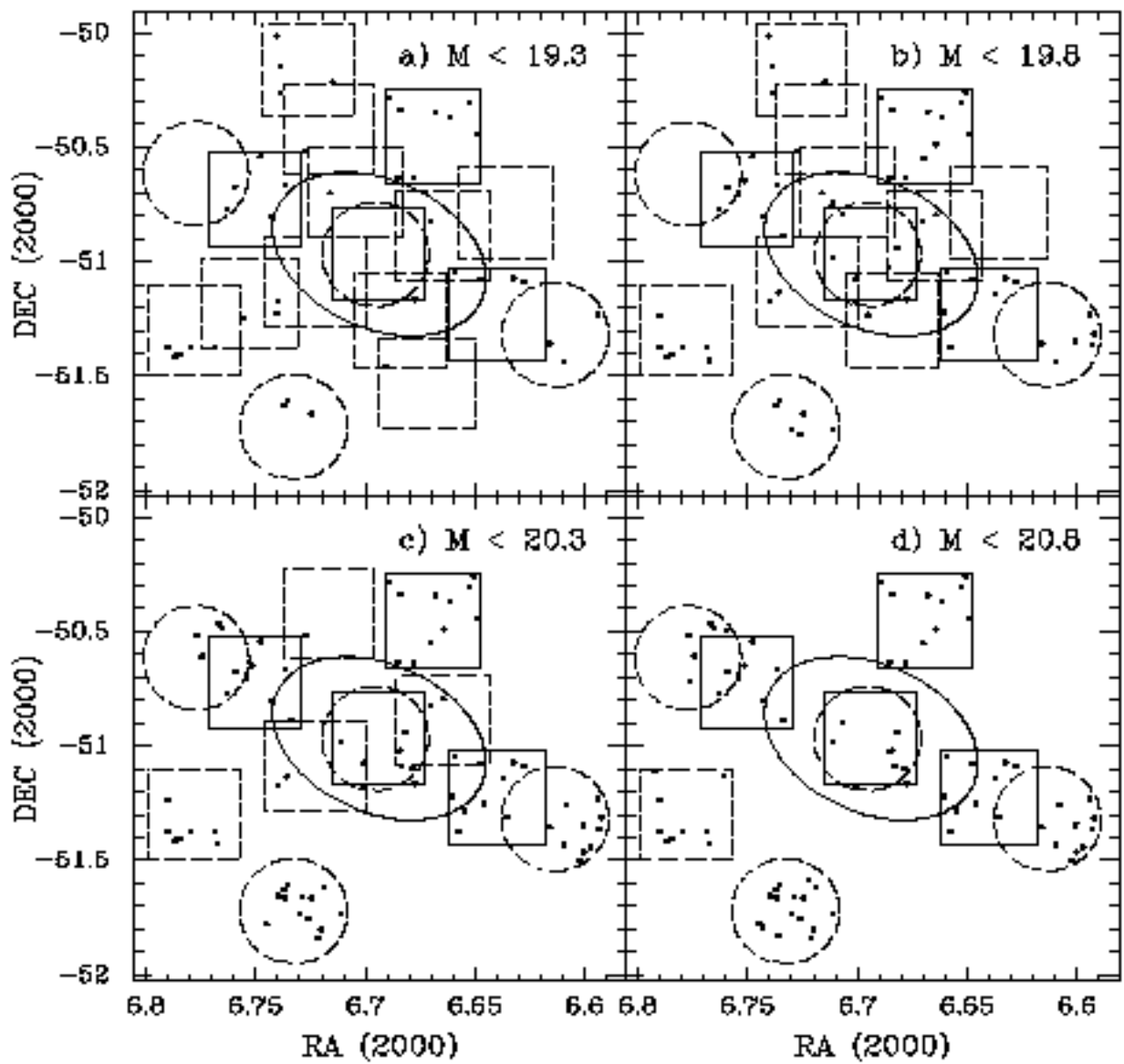
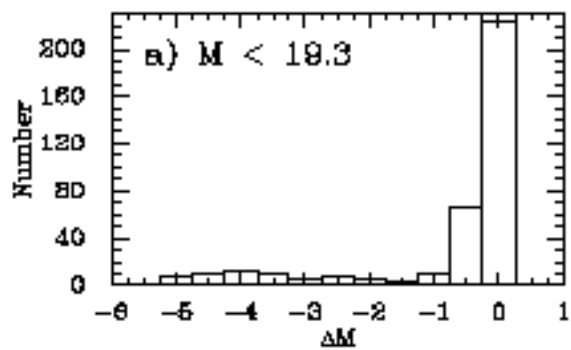
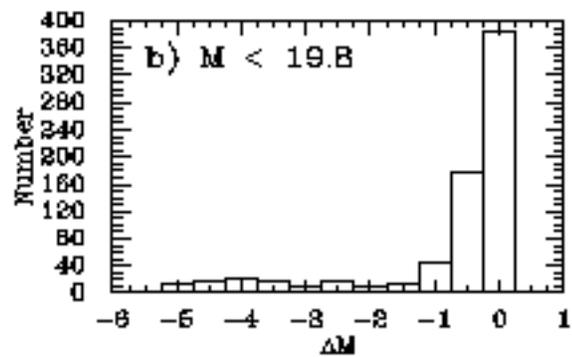
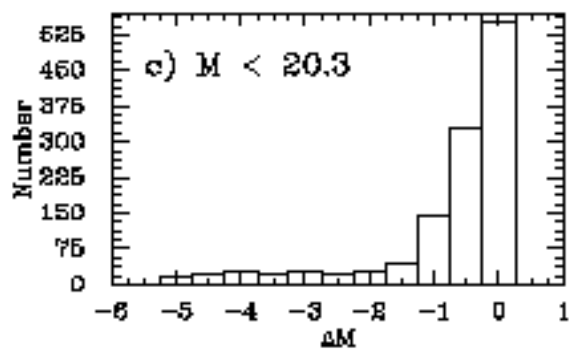
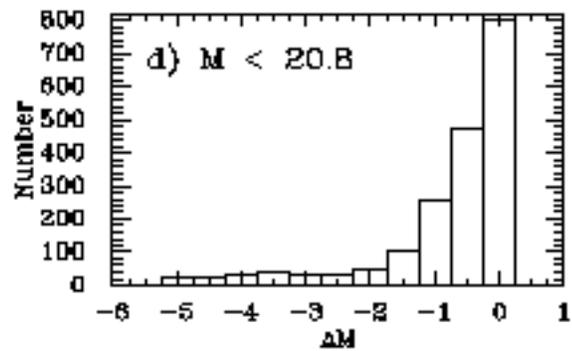
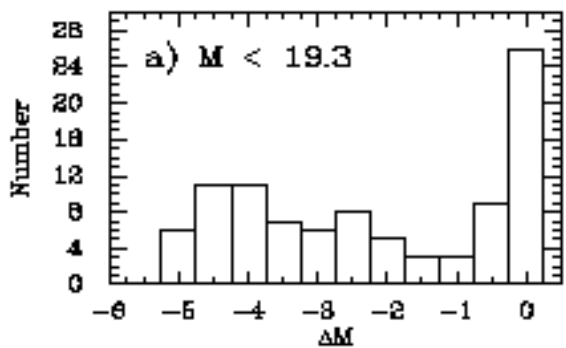
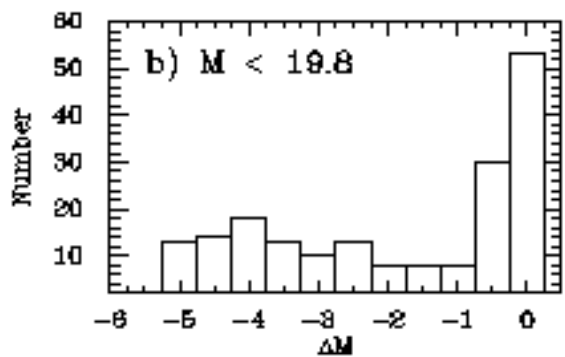
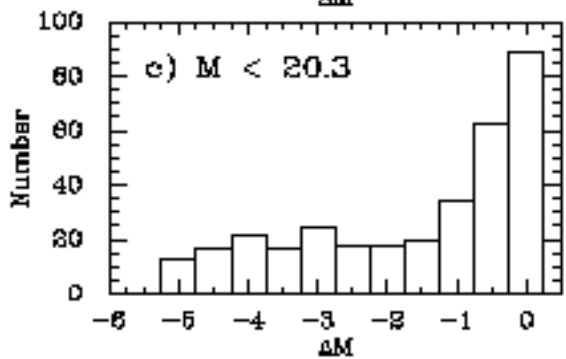
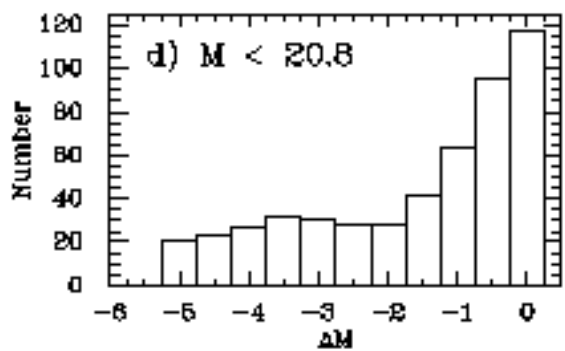
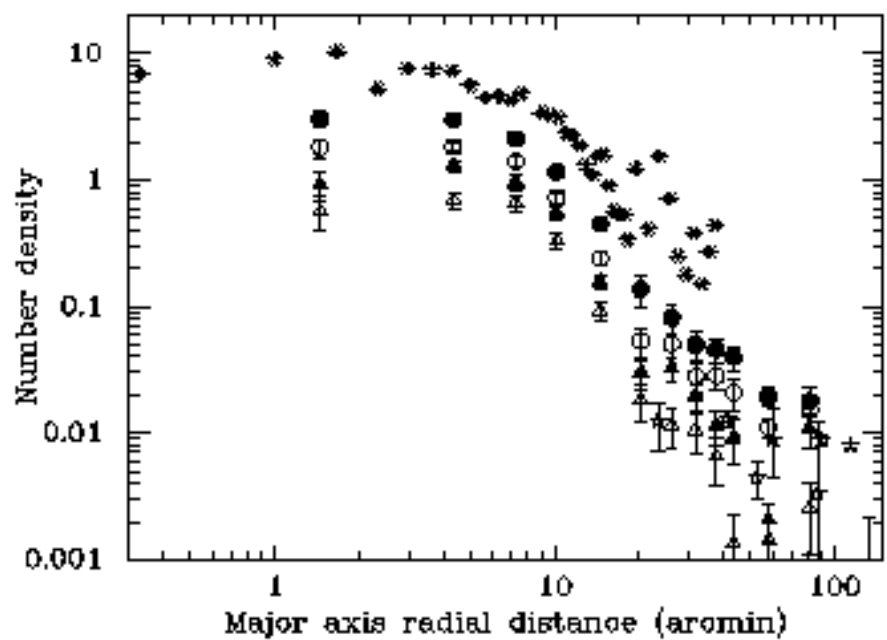
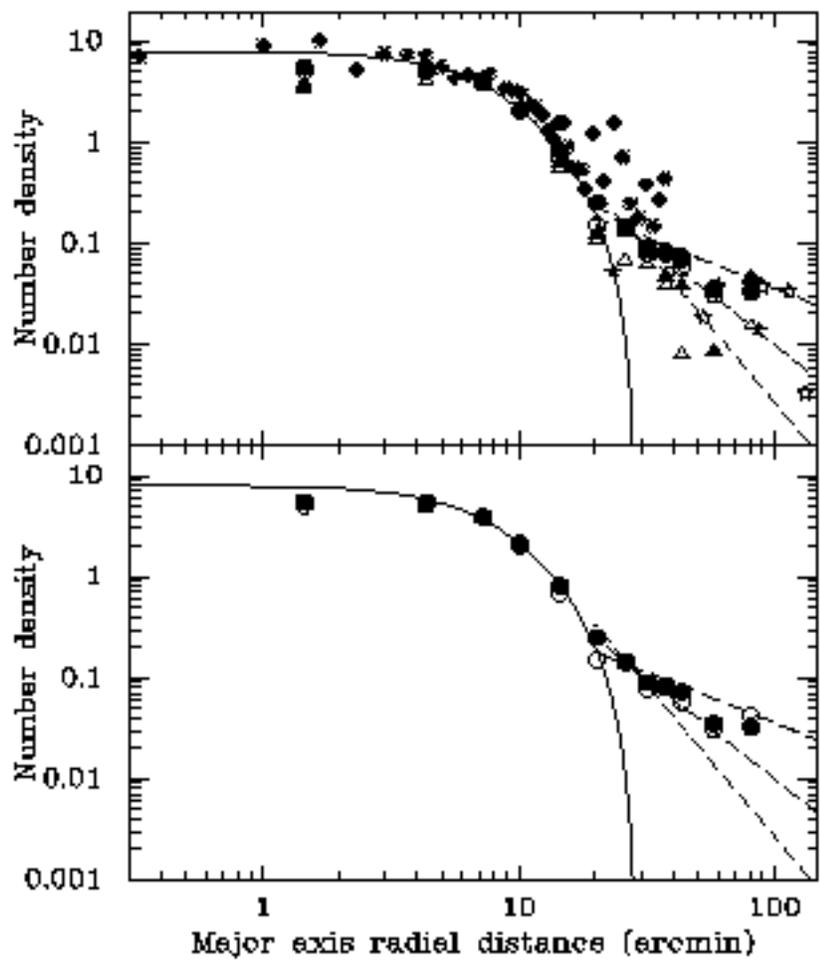
\end{document}